\documentstyle[11pt,aaspp4]{article}

\renewcommand{\sun}{\odot}
\newcommand{\degr}{$^\circ$}

\newcommand{\h}{H\scriptsize II \normalsize}
\newcommand{\hi}{H\scriptsize I \normalsize}
\def\etal{et al.}
\def\kms{km~s$^{-1}$}

\begin{document}            

\title{THE HII REGION KR 140: SPONTANEOUS FORMATION OF A HIGH MASS
STAR}

\author{D. R. Ballantyne\altaffilmark{1}, C. R. Kerton\altaffilmark{2}
 and P. G. Martin}
\affil{Canadian Institute for Theoretical Astrophysics, University of
Toronto, Toronto, ON, Canada~M5S~3H8; ballanty, kerton and
pgmartin@cita.utoronto.ca}

\altaffiltext{1}{Current address: Institute of Astronomy, Madingley Road,
 Cambridge, United Kingdom CB3 0HA}

\altaffiltext{2}{Current address: Dominion Radio Astrophysical Observatory,
PO Box 248, Penticton, B.C., Canada V2A 6K3}

\begin{abstract}

We have used a multiwavelength data set from the Canadian Galactic Plane
Survey (CGPS) to study the Galactic \h region KR~140, both on the scale
of the nebula itself and in the context of the star forming activity in
the nearby W3/W4/W5 complex of molecular clouds and \h regions.  From
both radio and infrared data we have found a covering factor of about
0.5 for KR~140 and we interpret the nebula as a bowl-shaped region
viewed close to face on. Extinction measurements place the region on the
near side of its parent molecular cloud.  The nebula is kept ionized by
one O8.5~V(e) star, VES~735, which is less than a few million years old.
CO data show that VES~735 has disrupted much of the original molecular
cloud for which the estimated mass and density are about
5000~$M_{\odot}$ and 100~cm$^{-3}$, respectively.  KR~140 is isolated
from the nearest star forming activity, in W3.  Our data suggest that
KR~140 is an example of spontaneous (i.e., non-triggered) formation of,
unusually, a high mass star.

\end{abstract}


\section{INTRODUCTION}
\label{sec:intro}

Massive OB stars are almost always found in clusters.  In fact,
accumulating observational evidence suggests that most stars, regardless
of mass, actually form as members of some kind of group, cluster, or
association.  While there has been a lot of work to understand the
processes involved in the formation of a single star (e.g., Shu, Adams,
\& Lizano 1987), theories of cluster formation are still in their
infancy (see the recent reviews by Elmegreen \etal\ 2000, and Clarke,
Bonnell \& Hillenbrand 2000).  Nevertheless, many important issues have
already been identified.  Foremost among these is the question of
whether the formation of clusters, particularly ones with OB stars, are
triggered by an external agent (Elmegreen 1992) such as expanding \h
regions (Elmegreen \& Lada 1977), or colliding molecular clouds (Loren
1976, 1977; Scoville, Sanders \& Clemens 1986; Usami, Hanawa, \&
Fujimoto 1995).  Despite triggered or sequential star formation being
theoretically and intuitively appealing, a major problem is to determine
unambiguously a cause and effect relationship, because of the long time
scales (and time lags) of the processes involved.  There is
observational evidence for regions that have had triggered star
formation, both on large scales (e.g., IC~1396, Patel \etal\ 1998) and
on small scales (e.g., IC~1805, Heyer \etal\ 1996).  However, there are
other young star forming regions, like Taurus, where no evidence of a
trigger can been found.  These are often in modest-sized molecular
clouds and contain only lower mass stars.

The Perseus Arm star forming regions W3/W4/W5 (Westerhout 1958) have
been studied extensively over the last twenty years (e.g., Lada \etal\
1978; Braunsfurth 1983; Digel \etal\ 1996; Normandeau, Taylor \& Dewdney
1997; Heyer \& Terebey 1998).  They are often considered to be the
archetypical examples of how the formation of massive star clusters can
be triggered by the influence of other nearby clusters.  For example, W3
is thought to be have been triggered by the expansion of W4 (Dickel
1980; Thronson, Lada, \& Hewagama 1985; van der Werf \& Goss 1990) and
there is evidence that the expansion of W5 is also triggering star
formation (Vall\'{e}e, Hughes, \& Viner 1979; Wilking
\etal\ 1984).

In new high resolution multiwavelength (radio and mid-infrared) data of
the W3/W4/W5 complex from the Canadian Galactic Plane Survey (CGPS;
English \etal\ 1998) we have identified a star forming region containing
a single O star.  Figure~\ref{fig:1420all} shows a 1420~MHz continuum
image from the CGPS pilot project (Normandeau \etal\ 1997).  The circled
area is the \h region in question, KR~140 ($l=133.425^{\circ}$,
$b=0.055^{\circ}$; Kallas \& Reich 1980).  This region appears to be
completely separate from the vigorous star formation going on nearby in
W3, although it is in the same Perseus arm molecular complex (see
context in CO images in Heyer \etal\ 1998).  What is unusual is that
this massive star seems to have been formed spontaneously.

In this paper we present and analyze the multiwavelength data on KR~140
in order to quantify the properties of this region of spontaneous
massive star formation.  As described in \S~\ref{sec:data} these data
have sufficient resolution (1$^{\prime}$) to resolve this \h region for
the first time and reveal a fairly symmetrical structure.  Based on
first impressions, we thought that KR~140 might prove to be a
``textbook'' spherical \h region.  Instead, we find that KR~140 is likely a
bowl-shaped \h region (an example of a blister geometry: Israel 1978,
Yorke \etal\ 1989; \S~\ref{sec:blist} and \S~\ref{subsub:ion},
\S~\ref{sub:irmod}).  We have found that the \h region is kept ionized
by an O8.5V(e) star, VES~735, and is at a distance of 2.3$\pm$0.3~kpc
from the Sun (Kerton \etal\ 1999).

We analyze the ionized component of KR~140 in \S~\ref{sec:kr} and make
various estimates of the age in \S~\ref{sec:age}.  The dust, molecular,
and atomic components of KR~140 are examined in \S~\ref{sec:ir},
\S~\ref{sec:mol-cld}, and \S~\ref{sec:hi}, respectively.  In
\S~\ref{sec:discuss} we discuss star formation in KR~140 in the context
of the Perseus arm and the possible accompanying cluster.

\section{THE MULTIWAVELENGTH DATA}
\label{sec:data}

KR~140 was initially cataloged in a 1420~MHz radio continuum survey of
the northern Galactic plane with the Effelsberg 100~m telescope by
Kallas \& Reich (1980). At the available resolution of 9$^{\prime}$
KR~140 was barely resolved, with a reported diameter of about
11$^{\prime}$.  Although KR~140 was measured in other subsequent
single-dish surveys (\S~\ref{sub:prop}) it had never been examined with
a radio interferometer to provide the angular resolution necessary for
the present study.

The data analyzed here include 1420~MHz ($\lambda$ 21~cm) and 408~MHz
($\lambda$ 74~cm) continuum images from the Dominion Radio Astrophysical
Observatory (DRAO) Synthesis Telescope (Roger \etal\ 1973; Veidt \etal\
1985).  An \hi 21-cm line data cube is also available, with spectral
resolution 2.64~\kms\ and channel spacing 1.65~\kms.  These DRAO data
were obtained during the pilot project of the CGPS (Normandeau \etal\
1997), and made into a full 8$^{\circ} \times 5^{\circ}$ mosaic of the
W3/W4/W5 star forming regions.

We also made use of a CO ($J=1 \rightarrow 0$, 115 GHz, $\lambda$ 2.60
mm, spectral resolution 0.98 \kms ) data cube from the Five Colleges
Radio Astronomy Observatory (FCRAO) obtained in the complementary survey
described by Heyer \etal\ (1998).  To examine the dust components we
processed Infrared Astronomical Satellite (IRAS) data to make HIRES
mosaics at 12, 25, 60, and 100~$\mu$m.  Independent processing of the
entire Galactic plane data at 60 and 100~$\mu$m has been released as the
IRAS Galaxy Atlas (IGA, Cao \etal\ 1997) and we have completed a
complementary project at 12 and 25~$\mu$m (Mid-Infrared Galaxy Atlas,
MIGA, Kerton \& Martin 2000).

An important feature for subsequent analysis with these data sets is the
common relatively high angular resolution achieved.  The DRAO synthesis
telescope images have a resolution of $1^\prime \times 1.14^\prime$ at
1420 MHz (the 408~MHz DRAO data have proportionately 3.5 times lower
resolution), and the CO images are at a resolution of
50$^{\prime\prime}$ (beam sampled).  The HIRES images have non-circular
beams of size about 2$^\prime$ at 100~$\mu$m, 1$^\prime$ at 60~$\mu$m,
and somewhat less at 12 and 25 $\mu$m.  For detailed intercomparison,
pairs or groups of images have been convolved to the same beam shape.

\section{A MODEL FOR THE 3-DIMENSIONAL MORPHOLOGY}
\label{sec:blist}

Analysis of an \h region benefits from a knowledge of its
three-dimensional structure, but often the observed two-dimensional
morphology is too complex to interpret.  Fortunately, this is not the
case for KR~140.  Figure~\ref{fig:irall} shows the four HIRES images
of KR~140 at 12, 25, 60, and 100~$\micron$.  As discussed in
\S~\ref{sec:ir}, the 100~$\micron$ image best describes the dust
distribution in KR~140.  The 100~$\micron$ HIRES image
(Figure~\ref{fig:irall}{\em d}) shows that most of the dust in the
nebula has been swept into a shell-like structure and that we are
observing a region which is close to being circularly (axially)
symmetric.  In the 1420~MHz image (Figure~\ref{fig:kr140}) KR~140 is
also seen to be fairly circular with a central depression.  As
highlighted by the contour map, there are three high intensity areas --
two ``eyes'' and a ``mouth''.  However, where the ``nose'' would be is a
lower intensity area.  The simplest interpretation appears to be a
three-dimensional structure with a central hole (\S~\ref{sub:otherage}).
We have identified VES~735 as the exciting star of this \h region
(Kerton \etal\ 1999); it lies projected close to this lower intensity
area of \h emission, though not coincident with it.

Since we have also classified the exciting star, we can draw further
conclusions about the geometrical structure of KR~140 from a global
analysis (\S~5 in Kerton \etal\ 1999).  Even after taking into account
the emission from dust that was not observed by IRAS, we find that the
total luminosity of warm dust in the nebula is far less then the
bolometric luminosity of the ionizing star VES~735 and that the covering
factor of the dust is only about 0.4 -- 0.5 (see \S~\ref{sub:irmod}).
Furthermore, the total radio flux at 1420~MHz is lower than expected for
an ionization bounded nebula surrounding VES~735, with a similar implied
covering factor (\S~\ref{subsub:ion}).  This together with geometrical
information to be discussed (e.g., \S~\ref{sub:s&s}) implies that KR~140
is a bowl-shaped \h region (cf.\ Roger \& Irwin 1982), rather
than a classical Str\"omgren sphere, a not unexpected morphology when
the O star is close to the edge of its parent molecular cloud (Yorke
\etal\ 1983). This model is perhaps the simplest geometry that is
consistent with the derived covering factors, but our data cannot rule
out other, more complicated geometries, such as a broken shell or
shredded and dissipated Str\"omgren sphere. 
Observations of a champagne flow (Tenorio-Tagle 1979), perhaps via
Fabry-Perot imaging, could help establish that KR~140 is truly a blister
region.

The circular symmetry observed in the dust shell (and in the ionized
gas) implies that we are observing KR~140 almost face on (i.e., the
opening of the bowl is oriented almost along the line-of-sight). For
example, see the simulated radio maps of the R2 model by Yorke \etal\
(1983).  The CO data and extinction measurements to VES~735 both suggest
that the molecular cloud is behind the star, with the inferred opening
toward us (\S~\ref{sub:co}).

\section{KR~140 FROM THE PERSPECTIVE OF THE RADIO CONTINUUM}
\label{sec:kr}

\subsection{Size and Structure}
\label{sub:s&s}

The basic radio morphology appears to be that of a limb-brightened
hemispherical shell (most of the free-free emission would be from the
dense and thin ionization-bounded zone).  The last contour shown (at
$T_b$ = 5.0~K) in Figure~\ref{fig:kr140} has a diameter of 8.5$^\prime$
which for a distance $d = 2.3$~kpc corresponds to a physical diameter of
about 5.7~pc.  It is expected that neutral material could extend well
beyond the \h region, and indeed there is both dust (\S~\ref{sub:t&d})
and CO gas (\S~\ref{sub:co}) around KR~140.  The free-free surface
brightness falls off most quickly (the radio contours are more tightly
spaced) in the west-northwest part of the nebula.  A sharper ionization
front would occur in more dense (pre-existing) material.

Other internal detail might be interpreted as variations in distance
between the star and the background ionization front, with 
lower surface brightness features originating from sectors further from
the star; such an interpretation of radio surface brightness has been
used to construct a topological map of the ionized surface of the Orion
molecular cloud behind the Trapezium (Wen \& O'Dell 1995).  If we use
the 1420~MHz surface brightness at the pixels near the projected
position of VES~735 and the known properties of this O8.5V exciting
star, the distance that the star must be from the back of the nebula
turns out to be only 2.14~pc.  Together with the above estimate of the
diameter across the line of sight, this is consistent with the idea that
KR~140 is similar to a hollowed-out hemispherical bowl.

\subsection{Physical Properties}
\label{sub:prop}

In this section we use the 1420~MHz data to derive a number of physical
properties of the nebula. For ease of reference,
Table~\ref{tbl:physprop} summarizes these properties, some of the
properties of the exciting star, and other quantities that we later
derive from the infrared and CO data.
  
\subsubsection{Emission Measure, Radio Flux, Electron Density, and
Ionized Mass}
\label{subsub:em}

The KR~140 emission is optically thin at 1420~MHz with a peak brightness
temperature about 10~K.  For an optically-thin nebula, the specific
intensity (surface brightness) $I_\nu$ and brightness temperature
$T_{b\nu}$ for thermal radiation are (Osterbrock 1989):
\begin{equation}
\label{eq:ite}
I_\nu \equiv 2 \nu^2 k T_{b\nu}/c^2 = j_\nu \int n_i n_e ds \equiv 
j_{\nu}E,
\end{equation}
where $j_\nu$ is the free-free emissivity, $n_i$ is the ion density,
$n_e$ is the electron density, $ds$ measures distance along the line of
sight, and the integral $E$ is called the emission measure.  For
$j_{\nu}$ of a pure hydrogen nebula (a good approximation; see
\S~\ref{subsub:ion} for models which include ISM abundances),
$E = 5.77 \times 10^2 T_{b\nu} (T_e/7500 {\rm K})^{0.35} (\nu/1420 {\rm
MHz})^{2.1}$~cm$^{-6}$~pc.
The peak emission measure in the map is about 6000 cm$^{-6}$~pc and
after background subtraction $\langle E\rangle = 2050$~cm$^{-6}$~pc.
After subtracting the local Galactic background we find the total flux
($F_{\nu} = \int I_{\nu} d\Omega$) to be $2.35 \pm 0.05$~Jy at
1420~MHz. This is in good agreement with the value $2.3 \pm 0.2$~Jy
reported by Kallas \& Reich (1980).  Becker, White, \& Edwards (1991)
detected this \h region in a 6~cm survey with the NRAO 91~m.  Their
source, designated [BWE91]0216+6053, had $F_{\nu}=2.5$~Jy.  Using the
same telescope and frequency Taylor \& Gregory (1983) and Gregory \&
Taylor (1986) record this source as GT~0216+608 in their survey; in
their analysis they treated all sources as point sources and so find
systematically low fluxes for extended nebulae like this (648~mJy for
KR~140).

In the optically thin limit the theoretical expectation for the spectral
index for thermal radiation (defined as $I_\nu \propto \nu^{\alpha}$) is
$\alpha \approx -0.1$ (Oster 1961; Gordon 1988). The spectral index was
measured between 1420 and 408~MHz.  First, the distribution across the
nebula was determined using L. Higgs' ``specmap'' routine (Zhang \&
Higgs 1997), which evaluates the spectral index for the spatially
variable component of the emission (this approach has the advantage of
circumventing uncertainties in background subtraction in each image);
for this procedure, we convolved the 1420~MHz image to the lower
resolution of the 408~MHz image.  The typical value of $\alpha$ is $0.02
\pm 0.08$.  We confirmed this by measuring the total
background-subtracted flux at 408~MHz as well.  Considering there is a
10\% uncertainty in the flux calibration in the 408~MHz pilot project
data at the position of KR~140 (Taylor 1999), the derived spectral index
is in satisfactory agreement with theory.

Depending on details of the geometry, the appropriate pathlength for
estimating the density would be of order the observed radius; with this
choice we find an rms $n_e = 27$~cm$^{-3}$.  The mass of
radio-observable ionized gas (allowing for He) is
163~(27~cm$^{-3}$/$n_e$)~$M_\odot$.  For comparison, the exciting star
of KR~140 has a mass of 25~$M_{\odot}$ (Kerton \etal\ 1999).

\subsubsection{Number of Ionizing Photons}
\label{subsub:ion}

In ionization equilibrium the number of ionizations (where H is the
predominant species) that occur each second in the nebula is equal to
the number of (H$^+$) recombinations each second, both locally and
globally.  An ionization-bounded nebula occurs when there is sufficient 
material to intercept all of the ionizing photons ($\lambda \leq 912$
\AA), $Q(H^0)$, emitted by the star per second (Osterbrock
1989). Formally, the spatially-integrated radio emission of an
ionization-bounded nebula is an effective calorimeter for $Q(H^0)$:
\begin{equation}
\label{eq:iq}
fQ(H^0) = \int  n_e n_{\rm H^+} \alpha_B dV = \alpha_B F_{\nu} d^2/j_{\nu}
\end{equation} 
where $\alpha_B$ is the case B hydrogen recombination coefficient of
H$^+$ ($\approx 3.3 \times 10^{-13}$~cm$^3$ s$^{-1}$ at 7500~K; Storey
\& Hummer 1995), $dV$ is the volume element, and we have used $dV = ds
d\Omega \times d^2$ and Equation (\ref{eq:ite}) to find the right hand
side.  Here $f$ is the product of a number of correction factors
including $f_{ci}$, the all-important covering factor by an ionization
front: if the star is only partially surrounded by gas then a fraction
of the ionizing photons will escape and a lower $F_\nu$ will result.

Equation~\ref{eq:iq} is based on recombination and emissivity for a pure
H nebula at a fixed temperature and so $f$ is made up of a number of
factors which modify the expected radio continuum flux for a given
$Q(H^{0})$.  We used the spectral synthesis code Cloudy (Ferland \etal\
1998) to analyze the observations of KR~140 in the radio, particularly
the correction factors which comprise $f$.  We constructed a number of toy 
models for differential comparisons, using a 35,500~K blackbody with 
$\log[Q(H^0)] = 48.06$ and $n_e = 50$ cm$^{-3}$; see Table~\ref{tbl:radmod}.  
The differential results are appropriate for any low-density \h region 
heated by a late O-type star.

The temperature of the nebula enters in the equation through the
$\alpha_B$ and $j_\nu$ factors, although the effect on the ratio is very
small.  Comparing models 1 and 5 we estimate $f_T = 0.99$.

The effect of the addition of other elements to the nebula is to
increase the total radio flux.  This effect is primarily due to the
presence of He$^+$ increasing the effective free-free emissivity (He 
has a tiny effect on the amount of ionization of H; e.g., Osterbrock 
1989).  Comparing models 1 and 4, or models 2 and 5 we
estimate that $f_{He} = 1.07$.

Dust competes with the gas for the absorption of ionizing photons and
thus, when present in the nebula, will reduce the observed radio $F_\nu$
for a given $Q(H^{0})$.  Comparing models 1 and 2 (with grains) to model
3 (without grains) we derive $f_{dust} = 0.78$ for the typical ISM
grains used in model 1 and $f_{dust} = 0.93$ for dark-cloud dust,
similar to the dust found in Orion, used in model 2.  The dark-cloud
dust is not as efficient as typical ISM dust in absorbing
short-wavelength photons and so has less of an effect on the emergent
radio flux.

The combination of these factors, except for the covering factor, leads
to a factor of $f_{model} = 0.91\pm0.08$ depending upon the grain
composition (i.e., $f=0.91f_{ci}$).

In our study of the exciting star VES~735 (Kerton \etal\ 1999) we used
the measured radio flux at 1420~MHz ($2.35 \pm 0.05$~Jy) to calculate
$\log[fQ(H^0)]=48.05 \pm 0.11$, and then estimated $f_{ci} \approx
0.4-0.5$ based on the known spectral type of VES~735 (O8.5 V(e);
$\log[Q(H^0)]\sim 48.45$, Panagia 1973). We have not attempted to put
any formal error estimate on the covering factor, but given the
uncertainties involved, both in the stellar properties and in estimating
the various $f$ factors, this result is certainly consistent with the idea 
of KR~140 being an open bowl-shaped region.  As demonstrated in
\S~\ref{sub:irmod}, an analysis of the IR data leads to the same
conclusion independently.

\section{THE AGE OF KR~140}
\label{sec:age}

The approach we have adopted here is first to date the \h region using
data we have on the exciting star VES 735.  With that age in mind, we
then investigate the dynamics of the region, the goal being to show that
certain scenarios for the evolution of KR 140, such as it being a
blister region, are at least consistent with the age suggested from the
exciting star.  This approach is similar to that adopted by Dorland et
al.\ (1986) in their study of the Rosette Nebula.

\subsection{Stellar Content}
\label{sub:stellar}

The idea of using the stellar content of a \h region to measure the age
of the nebulosity was first attempted by Hjellming (1968).  Basically
one plots the evolutionary tracks of stars with various masses in the
$\log(L/L_{\odot})$ vs. $\log T_{eff}$ plane.  One obtains $T_{eff}$
from the spectral type of the exciting star and $\log(L/L_{\odot})$ from
the radio flux (much like $Q(H^0)$) and then the position in the plane
determines an age for the star and thus the \h region.  Clearly the
effectiveness of this technique depends strongly on the quality of the
calibration between theoretical and observational quantities as well as
the quality of stellar models, which have vastly improved in the thirty
years since this technique was introduced.  In the early work the
primary result was to indicate whether the \h regions were ionization or
density bounded, and that many \h regions required additional,
unobserved, sources of ionization.

Here we follow a slightly different technique compared to Hjellming
(1968) in order to avoid any uncertainties associated with the covering
factor and structure of the nebula.  We instead use the absolute
magnitude ($M_V$) as a measure of $\log(L/L_{\odot})$, which is possible
because we have a good estimate of the distance to the star and the
extinction along the line of sight (Kerton \etal\ 1999).
Figure~\ref{fig:age} plots the stellar evolution models of Schaerer \&
de Koter (1997) for a 20, 25 and 40 $M_\sun$ star along with the value
determined for VES 735.  The observed values for VES 735 are consistent
with a 25 $M_\sun$ star with an age of a few million years away from the
ZAMS (see Figure \ref{fig:age}), where the age would be only of order
$10^5$ years.  Of course, the spectral type alone gives us a simple
upper limit to the age of KR~140: for an O8.5 V star the main sequence
lifetime is $\sim 6 \times 10^{6}$ (Chieffi \etal\ 1998).

\subsection{Dynamical Models: Spherically-symmetric}
\label{sub:dynamics}

\subsubsection{Str\"omgren Sphere}
\label{sub2:sphere}

The simplest description of a \h region is that of the formation and
expansion of a ionized ball of pure hydrogen at constant temperature in
an uniform medium of constant density (Str\"omgren 1939).  We summarize
this only as a point of departure and contrast.  The evolution of a
Str\"omgren sphere starts with a formation phase where the O star
ionizes a region of space around it to the radius ($R_s$) given by:
\begin{equation}
R_s=  \left( \frac{3Q(H^0)}{4\pi n_e^2 \alpha_B} \right)^{1/3}.
\end{equation}
The initial rapid expansion to a radius of $r_i$ occurs in time $t =
(-1/n_e \alpha_B)\ln(1-(r_i/R_s)^3$) (valid to $r_i/R_s \sim 0.98$,
Osterbrock 1989).  The time to create the initial Str\"omgren sphere is
about $10^5 / n_e$ (cm$^{-3}$)~y, instantaneous compared to other
timescales.  For KR~140, the observed radius ($R_{obs}$) is 2.85 pc.
Taking $\log[Q(H^0)] = 48.45$ for an O8.5 V(e) star like VES 735
(Panagia 1973) and $n_{H_2} = 100$ cm$^{-3}$ for the original molecular
cloud, $R_s = 1.2$ pc. The initial \h region is overpressured compared
to the surroundings and will expand into a uniform medium according to a
$t^{4/7}$ law:
\begin{equation}
r_i/r_o = \left(1 + \frac{7C_{II}t}{4r_o}\right)^{4/7},
\end{equation}
where $C_{II}$ is the sound speed in the ionized medium, $r_o$ is the
initial radius and $r_i$ is the radius at time $t$.  With the densities
quoted above the \h region will evolve from $R_s$ to $R_{obs}$ on a
timescale of 10$^5$ years, which is improbably short. Increasing the
initial density makes $R_s$ smaller and forces the pressure expansion
stage to be longer. A value of $n_{H_2} = 500$ cm$^{-3}$ will push the
timescale to 10$^6$ years.  However, this value is not consistent with
our observations of the molecular material (\S~\ref{sec:mol-cld}), so
the Str\"omgren sphere is not an appropriate dynamical model for KR~140.

\subsubsection{Stellar Winds}
\label{sub:otherage}

In the radio image a local minimum is evident near the position of the
central star (see Figure~1 in Kerton \etal\ 1999).  One interpretation
is that this is a wind blown bubble around the O8.5 V(e) star.  The
apparent radius of the central hole is 1.4 pc.  We used the model of
Castor \etal\ (1975) for the size of a circumstellar shell:
\begin{equation}
R = 28 \left( \frac{\dot{M}_6 V_{2000}^2}{n} \right)^{1/5} t_{6}^{3/5}
\mathrm{pc}
\label{eq:shell}
\end{equation}
where $R$ is the shell's radius (pc), $\dot{M}_6$ is the mass loss rate
($10^{-6}$~$M_\sun$~yr$^{-1}$), $V_{2000}$ is the wind velocity (2000 km
s$^{-1}$) $n$ is the gas density (cm$^{-3}$), and $t_6$ is the time
($10^6$ years).  Stellar wind properties of the B2 model of Schaerer \&
de Koter (1997) were used.  For a wide range of densities we obtain
timescales on the order of only 10$^4$ years. This is far too low a
timescale and clearly this model is not an appropriate description of
VES 735.  Either the mass loss rate adopted is much too high, or the
assumption of spherical symmetry is not valid.

\subsection{Dynamical Models: Open Geometry}
\label{sub2:blister}

Evidence summarized thus far (\S~\ref{subsub:ion} and \S~\ref{sub:irmod})
indicates an open geometry with a covering factor of about 0.5 for both
dust and ionized material.  In this case the high pressure in the
interior of the initial Str\"omgren sphere or stellar wind bubble is
relieved, slowing down the expansion.  Models have been developed to
investigate the evolution of an \h region at the edge of a molecular
cloud (see review of early models by Yorke (1986), and Comer\'{o}n
(1997) for a recent example). While complex numerical modeling is
possible, Franco et al.\ (1994) have shown that the expansion of the
ionization bounded side of such a blister \h region is well described by
\begin{equation}
\label{eq:tbf}
r_i/r_o = \left(1 + \frac{5C_{II}t}{2r_o}\right)^{2/5}.
\end{equation}
This is based upon mass conservation between material being ejected in
the flow and molecular material being eroded off the cloud.  One can
envisage the evolution of a blister as consisting of three stages: the
initial rapid formation stage, a pressure driven expansion stage, and
finally a blow-out stage.  The relative length of time of the latter two
stages depends upon the distance of the star from the edge of the cloud
and the density structure of the cloud.  One very important point is
that an O star very close ($\sim 1 R_s$) to the edge of a cloud will
very quickly develop a covering factor of $\sim 0.5$ in order 10$^5$
years and will maintain this covering factor over the lifetime of the O
star (Yorke \etal\ 1983, 1989).  We assume that the star formed very
close to the edge of the cloud, thus ignoring the pressure driven
expansion stage.  Using Equation~\ref{eq:tbf}, the age of the region is
of order a million years when $R_s = 0.64$ pc.  For the stellar
properties of VES 735 this requires $n_{H_2} \sim 280$ cm$^{-3}$.  This
is somewhat encouraging as it does not require as vast a difference
between the properties of the observed molecular cloud and the putative
initial conditions.

\section{KR 140 IN THE INFRARED}
\label{sec:ir}

\subsection{Morphology and Emission Mechanisms}
\label{sub:irs&s}

Stars that are forming and evolving in the ISM interact with the
interstellar dust component by heating, redistributing, and possibly
destroying it.  The dust can be heated by at least three distinct
mechanisms: direct radiation from the central star, reprocessed
radiation from the ionized gas, and diffuse radiation from the
interstellar radiation field.  Radiation pressure from the star acts on
the dust in the ionized zone (Spitzer 1978) which causes the dust (and
gas) to be pushed away from the star.  Some forms of dust like
polycyclic aromatic hydrocarbons (PAHs) are destroyed in intense
ultraviolet fields.  A morphological study of the infrared emission from
KR~140 is therefore important to fully understand the energetics and
effects on the environment.

As mentioned in \S~\ref{sec:data}, IRAS scans of KR 140 at 12, 25, 60,
and 100 $\mu$m were processed by the HIRES software to generate maps of
about 1$^{\prime}$ resolution. The new beam shapes are somewhat
elliptical and so point sources will be visibly stretched; however, the
larger scale morphology of the observed dust emission from the nebula
will not be greatly affected by the asymmetric beam.

The intensity of the dust emission, $I_{dust}$, at a particular
frequency, $\nu$, from a distance increment $ds$ along the line of sight
has the form
\begin{equation}
\label{eq:idust}
I_{dust} = N_{dust} \pi a^2 B_{\nu}(T_{dust}) Q_{\nu}(a,T_{dust}) ds,
\end{equation}
where $N_{dust}$ is the number density of grains, $B_{\nu}(T_{dust})$ is
the Planck function at a temperature $T_{dust}$, $a$ is the radius of
the particles, and $Q_{\nu}(a,T_{dust})$ is the absorption efficiency
factor.

Figure~\ref{fig:irall} shows the four HIRES images of KR~140, with the
12 and 25 $\mu$m images convolved to the 1420 MHz resolution, and
overlaid by 1420 MHz contours (Figure~\ref{fig:irall}{\em a} \& {\em
b}). The 12 and 25 $\mu$m emission is spread well outside the radio
contours of KR~140 (this is taken up in \S~\ref{sub:25um}). Features
that are common to all four images in Figure~\ref{fig:irall} are the
bright arcs on either side of the nebula (easiest to see in
Figure~\ref{fig:irall}{\em c}) that extend outside the radio contours.
We interpret this as the limb-brightened warm dust shell around KR 140.
Note that there could be cooler dust further out around the KR~140
complex that will not have been detected in the IRAS bands.  We account
for the energetics of this cooler dust in our models of KR~140
(\S~\ref{sub:irmod}).

\subsection{Temperature and Column Density}
\label{sub:t&d}

Equation~\ref{eq:idust} can be used to calculate a mean dust temperature
for each pixel of IR emission.  Taking the intensity ratio for any two
frequencies, $\nu_1$ and $\nu_2$, yields:
\begin{equation}
\label{eq:iratio}
{I_{\nu_2} \over I_{\nu_1}} = \left(\nu_2 \over \nu_1 \right)^{3 + \beta} \left({e^{h\nu_1 / kT} -1 \over e^{h\nu_2 / kT} -1 }\right),
\end{equation}
where $T$ is the $T_{dust}$ of Eq.~\ref{eq:idust}, and the $\left(\nu_2
/ \nu_1 \right)^{3 + \beta}$ factor is from the frequency-dependent part
of $Q_{\nu}(a, T)$, where $\beta$ depends on the type of dust. The most
common components proposed for interstellar dust are silicate and
graphite (or some related carbonaceous material), which, at long
wavelengths, have $\beta \approx 2$. This value of $\beta$ is close to
what is generally observed in the ISM (Lagache \etal\ 1998).

The 60 $\mu$m and 100 $\mu$m data were used to calculate the dust
temperature map, as they are the bands where classical grain emission
dominates.  It is likely that non-equilibrium heating of very small dust
grains (VSGs) contributes some of the observed 60 $\mu$m flux and so the
derived temperatures are probably slight overestimates of the true grain
temperatures (Boulanger \etal\ 1988).  The images were brought to the
same resolution and background subtracted and the IPAC analysis program
`cttm' was used to compute a dust temperature map.  An
azmuthially-averaged radial cut of this map is shown in
Figure~\ref{fig:tdust}. The temperature distribution was sampled every
$1'$ in radius and every $10^{\circ}$ azimuthally. The temperatures in
the central region are around 31 K. Nearer the edge of the \h region,
the dust temperatures drop to around 28 K.  However, a line of sight
passing near the star also has cooler dust in the background and the
foreground, which would lower the apparent temperature observed.

A calculation of dust temperature from first principles was done as a
check on the empirical values output from `cttm'. In this calculation, a
single silicate or graphite dust grain with a radius of 0.1 $\mu$m (a
typical interstellar size, see Kim, Martin \& Hendry (1994)) was placed
at a distance of 3.0 pc from the center of the nebula, corresponding to
a dust grain within the KR~140 dust shell. The luminosity of the
O8.5V(e) exciting star is about 10$^5$~$L_{\odot}$ (Panagia 1973).
Making use of Planck-averaged absorption factors from Laor and Draine
(1993), we find that if $\tau_{UV} \approx 1$, the calculated dust
temperature is about 28 K, in agreement with the temperature deduced
empirically. Note, however, that this calculation did not take into
account the recombination-line photons emitted by the ionized
gas. Furthermore, some of the free-free photons and
collisionally-excited cooling lines are emitted not in the ultraviolet,
but in the optical (Osterbrock 1989), where the dust absorption
efficiency is somewhat lower.

The optical depth $\tau_\nu$ from dust is defined to be
\begin{equation}
\label{eq:tau}
\tau_\nu = \int N_{dust} \pi a^2 Q_{\nu}(a, T_{dust}) ds,
\end{equation}
and can be calculated by dividing $I_\nu$ (Equation~\ref{eq:idust}) by
the Planck function $B_\nu(T)$, assuming a constant $T$ along the line
of sight, adopted from the temperature map.  Figure~\ref{fig:tau100}
displays a dust optical depth map at 100 $\mu$m. The values range from
about 0.0004 in the middle of the nebula to about 0.002 in the bright
northwest rim. These values show that the dust is transparent to its own
100 $\mu$m emission. The minimum in the center of the nebula and the
ring-shaped appearance is most simply interpreted as limb brightening in
a thick shell of dust with the highest column density along the
northwest rim. The latter is the same region of the nebula where the
1420~MHz radio contours fall off most steeply (\S~\ref{sub:s&s}), and
where there is no CO emission (\S~\ref{sub:co}).

The optical depth in the ultraviolet $\tau_{UV}$ can be gauged using
extinction curves from the literature.  We are interested in the radial
as opposed to line of sight optical depth.  Judging the thickness of the
shell from one of the arcs in Figure~\ref{fig:tau100} gives a path
length of about 7 $\times 10^{18}$~cm. If we assume the dust is
associated with gas at a molecular density of 100~cm$^{-3}$, the
molecular column density of this gas is about 7~$\times
10^{20}$~cm$^{-2}$.  Using the extinction curves of Kim, Martin \&
Hendry (1994), we find that $\tau_{\nu} \approx 2.8$ at 1100~\AA\ (far
in the ultraviolet), whereas at 5500~\AA\ (in the optical), $\tau_{\nu}
\approx 0.6$.  The estimated optical depths show that the dust in the
arcs has a high enough radial optical depth to absorb most of the
incident ultraviolet photons.  However, since KR~140 has a covering
factor of 0.4 -- 0.5 (\S~\ref{subsub:ion} and \S~\ref{sub:irmod}), the
total infrared luminosity of the nebula will be correspondingly less
than the bolometric luminosity of VES~735 (\S~\ref{sub:irmod}).
 
\subsection{The 12 and 25~$\mu$m Emission}
\label{sub:25um}

It has been known for over a decade that there is excess emission within
the 12 $\mu$m IRAS passband (e.g., Boulanger, Baud \& van Albada
1985). Onaka \etal\ (1996) found that more than 70\% of the 12 $\mu$m
diffuse interstellar emission detected by IRAS is emitted in spectral
features attributed to polycyclic aromatic hydrocarbons (PAHs; L\'{e}ger
\& Puget 1984; Allamamdola, Tielens \& Barker 1985). The 12~$\mu$m image
of KR~140 is very instructive.  There is very little 12~$\mu$m flux
within the radio contours of KR~140, implying that PAHs are destroyed
there in the intense ultraviolet radiation field.  However, there is a
large amount of diffuse flux outside the radio contours, especially to
the west (Figure~\ref{fig:irall}\emph{a}). The PAH emission is a tracer
of the photodissociation region around a nebula (Giard \etal\ 1994;
Bregman \etal\ 1995; Fig.~\ref{fig:co12}).

Unique among the HIRES images of KR~140, the 25~$\mu$m image
(Fig.~\ref{fig:irall}b; Fig.~\ref{fig:ir25um}) shows a bright spot near
the center of the nebula, on a transition between a radio peak (the
``left eye'') and the deepest depression.  The IRAS Point Source Catalog
(Joint IRAS Science Working Group 1988) lists this feature as
IRAS~02165+6053, and it also has been identified with VES~735 (Bidelman
1988).  Figure~\ref{fig:ir25um} shows that IRAS~02165+6053 and VES~735 are
practically coincident.  Dust closer to the star would tend to
be warmer, contributing to the spot if not pushed away by radiation
pressure.  

The entire nebula looks hotter and more extended at 25~$\mu$m than it
would be for equilibrium emission from normal-sized grains.  Most of the
25~$\mu$m emission is probably contaminated by non-equilibrium emission
from very small grains (VSGs; Sellgren 1984) which have absorbed a UV
photon and have had their temperatures instantaneously rise to $\sim
10^3$~K.  Although these small grains make up a tiny fraction of the
mass in the dust distribution, they make up a good fraction of the
number distribution and absorb a significant fraction of the
near-ultraviolet radiation.  This VSG emission is therefore a major
source of uncertainty when analyzing the 25~$\mu$m
image.\footnote{Further evidence for contamination in the 25~$\mu$m IRAS
passband was presented by Cox (1990). He suggests that an iron oxide
emission line falls within this IRAS passband which would result in even
greater excess 25~$\mu$m emission.}  Note that the
``temperature-spiking'' phenomenon also occurs outside the ionized zone,
far from the exciting star.

\subsection{Infrared Models}
\label{sub:irmod}

\h regions are some of the most luminous objects in the Galaxy when
observed in the infrared, especially at wavelengths longer than 60
$\mu$m. In fact, if the dust shell covers 4$\pi$ steradians around the
exciting star, the infrared luminosity should be a good measure of the
star's bolometric luminosity. To account for the true extent of the
dust, we can define a covering factor of dust ($f_{cd}$).  Often it is
simply assumed that $L_{ir} = L_{bol}$; however, this is not correct
even for $f_{cd} = 1$.  Some of the radiation from the star and nebula
is at long enough wavelengths to avoid being absorbed by the dust.
Using Cloudy we found that for models with $f_{cd} = 1$,
$\log\left(L_{ir}/L_{bol}\right) \sim -0.1$.

To calculate the integrated infrared flux from KR 140, the background in
each of the four HIRES images was fitted and subtracted, and the total
flux from the nebula was measured in each band. In order to estimate
$L_{bol}$ using these data, we used Cloudy to simulate the emission from
large classical grains, roughly matching the fluxes at 60 and 100
$\mu$m.  This approach allows us to account for emission from grains
with a range of temperatures and thus unobserved emission at long
wavelengths.  First, we constructed models with $f_{cd} = 1$.  We find
$\log(L_{FIR}) = 37.90$.  Since the resulting spectrum misses most of
the observed 12 and 25 $\mu$m flux (which is caused by temperature
spiking of VSG's and the excitation of PAH molecules) we converted the
observed 12 and 25 $\mu$m fluxes to luminosities using tophat
approximations to the IRAS passbands (Emerson 1988) and added the
results to $L_{FIR}$.  With this addition we find $\log(L_{IR}) =
38.00$.  Correcting from $L_{ir}$ to $L_{bol}$ we find
$\log\left(L_{bol}/L_\odot\right) = 4.52$.  This is
significantly below what would be expected for any late O main sequence
star (e.g., Panagia 1973).  We interpret this low apparent $L_{bol}$ as
being due to a covering factor of 0.4 -- 0.5.

We know that VES 735 is an O8.5 V(e) star.  Figure~\ref{fig:cloudy2}
demonstrates that one can reproduce the observed $L_{ir}$ using the
appropriate stellar parameters and a covering factor of about 0.5.  A
single temperature (T = 28.25 K) $\nu^2B_\nu$ spectrum is shown for
comparison; note the wider model curve caused by the range of dust
temperatures contributing to the emission.

Using an alternative calibration of stellar parameters which includes 
wind-blanketed models (Kerton 1999), we find $f_{cd} \sim 0.5$ 
and $f_{ci} \sim 0.7$. These results are still consistent with a blister model 
for KR~140.

\section{KR~140 IN A MOLECULAR CLOUD}
\label{sec:mol-cld}

\subsection{CO Signature}
\label{sub:co} 

A slit spectrum of VES~735 allowed us to measure the radial velocity of
the nebular H$\alpha$ line within 30$^{\prime\prime}$ of VES~735 to be
$-46 \pm 2.1$~\kms with respect to the Local Standard of Rest [LSR]
(Kerton \etal\ 1999). The differential radial velocity between VES~735 and 
the nebular line was measured to be $+2.0\pm2.2$~\kms.

An expected effect of the evolution of the \h region is
photodissociation of molecular gas both inside the \h region and in the
immediate surrounding interstellar medium (ISM). The CO data traces the
molecular gas content of the ISM, so an examination of the CO data cube
ought to reveal a lack of emission near the ionized gas velocity of
KR~140.  Indeed, a distinct CO hole was found within the radio contours
of KR~140 over the velocity range $-$45.53~\kms to $-$47.16~\kms (LSR),
which corresponds to three channels of width 0.813~\kms\ in the CO cube.
At more negative velocities, CO emission from the parent molecular cloud
fills in the 1420~MHz contours.

Figure~\ref{fig:co-front} shows the sum of these three channels
overlayed with the 1420~MHz continuum contours. A well defined ring
structure is clearly seen.  Interestingly, the ring does not extend all
the way around the nebula: there is no CO emission in the
north-west. This is the same area where the 1420~MHz contours are
falling off more sharply (\S~\ref{sub:s&s}), there is a bright infrared
arc (\S~\ref{sub:irs&s}), and a bright \hi feature
(\S~\ref{sec:hi}). The relationship between 12 $\mu$m emission, which is
a good tracer of the PDR, and the CO emission is shown in
Figure~\ref{fig:co12}.  The north-west peak in the 12 $\mu$m emission
corresponds to the region where there is no CO emission. To the east, as
one moves away from the \h region the 12 $\mu$m peak occurs first
followed by the peak in the CO emission.

\subsection{Density and Mass of the Molecular Cloud}
\label{sub:co-mass}

To estimate the mass of the molecular cloud we integrated the CO cube
over the whole velocity range of the cloud ($-$45.5~\kms to $-$52.8~\kms),
and measured the total surface brightness of the molecular cloud.  We
can use the empirical $X$ factor to convert from CO surface brightness
to molecular hydrogen column density.  We have used $X=(1.9 \pm 0.5)
\times 10^{20}$~cm$^{-2}$~(K~\kms )$^{-1}$ (Strong \& Mattox 1996),
where the quoted uncertainty is to take into account the range of $X$ values
measured for a variety of different clouds, to obtain a mean column
density of N(H$_2$)$=2.2 \times 10^{21}$~cm$^{-2}$.  Assuming a path
length of 7~pc (the north-south radial extent) the density of the
molecular cloud is roughly $n_{H_2} = 100$~cm$^{-3}$, typical of a giant
molecular cloud (Blitz 1993).

By integrating the column density over the face of the cloud, we
estimate the mass of the cloud to be $(4.4 \pm 1.6) \times
10^{3}~M_{\odot}$.  This mass includes a He correction factor of 1.36,
and the error bar includes the distance uncertainty.  
The CO images show that the parent molecular cloud of KR~140 has been
greatly disrupted by the nebula and the exciting star VES~735 and so
this mass will be an underestimate of the cloud's initial mass.

\subsubsection{Estimating the Mass of the Original Molecular Cloud}
\label{subsub:co-origmass}

One way to estimate the mass of the original molecular cloud would be to
assume that KR~140 is indeed a blister \h region as our data
suggests. Yorke \etal\ (1989) have modeled blister \h regions with one
exciting O star and found that mass loss rate of material through the
blister is $3-5 \times 10^{-3}~M_{\odot}$~yr$^{-1}$.  However, these
authors modeled an O6 star in a cloud with $n_{H_2} = 500$~cm$^{-3}$,
which is not an accurate description of either VES~735 or its parental
cloud.  Unfortunately, Yorke \etal\ (1989) give no indication of how to
scale their mass loss rate for different values of $Q(H^0)$ or
$n_{H_2}$.  A analytic estimate of the amount of mass lost from a
blister \h region is given by Whitworth (1979). His Equation (41) has
the desirable property that it agrees with the results of Yorke \etal\
(1989) for the values used in their models; therefore, it might have the
correct scaling for $Q(H^0)$ and $n_{H_2}$. However, his calculations
were based on a cylindrical geometry which is not a very realistic model
for a more bowl-shaped spherically symmetric region such as KR~140.

In \S~\ref{sub2:blister}, we made use of the blister evolution formula
given by Franco \etal\ (1994). These authors considered a simple
symmetric blister region and were also able to estimate the cloud
evaporation rate:
\begin{equation}
\label{eq:mdot-franco}
\dot{M} \approx \pi R_S^2 m_p 2 n_{H_2} C_{II} \left ( 1 + {5 C_{II} t \over 2 R_S} \right )^{1/5},
\end{equation}
where $R_S$ is the Str\"{o}mgren radius and $m_p$ is the proton mass.
In \S~\ref{sub2:blister}, we found that an initial cloud density of
280~cm$^{-3}$ and $R_S= 0.64$~pc would match the observed radius of
KR~140 in a time of 1~Myr (consistent with the age of VES~735). Using
these values in the above equation gives a mass loss rate of $2.9 \times
10^{-4}~M_{\odot}$~yr$^{-1}$. Therefore, about 290~$M_{\odot}$ of
molecular material has been eroded from the cloud after 1~Myr.
Including the effects of dust lowers $R_S$, increases the age, and
lowers $\dot{M}$, resulting in a very similar amount of erosion 
(280~$M_{\odot}$).

Taking into account the mass lost through erosion and the present
ionized mass, we estimate the mass of the original molecular cloud in
which VES~735 formed to be about $4.9 \times 10^3~M_{\odot}$, so this
cloud would be classified as a dwarf molecular cloud (Elmegreen
1985). This seems to be a remarkably small cloud mass from which an O
star has formed.  Williams \& McKee (1997) performed a statistical
analysis of many nearby OB associations and molecular clouds, and found
that it was more likely that clouds with masses greater than
10$^5~M_{\odot}$ should form O stars than clouds with lower masses.
Other \h regions have been found with small molecular cloud masses
(Hunter \etal\ 1990), but O star formation with a cloud mass below
10$^4$ is considered rare (Elmegreen 1985).
Of course we have been attracted to this cloud because of the effect of
VES~735 and not to other parts of the W3 cloud complex (Heyer et al.\
1998) which have not formed O stars.

\subsection{Interpretation of the Velocity}
\label{sub:break}

Stepping through the velocity channels in the CO data cube near the
velocity of KR~140 shows that at higher (least negative) velocities
($\sim -40$~\kms) there is no emission, then the CO emission comes in at
the southern end of the \h region and spreads northward, before filling
in the 1420~MHz contours at a velocity of $-$48.78~\kms.  Perhaps the
best way to illustrate these data is by examining the cube in
velocity-latitude space.  Figure~\ref{fig:co-side} shows such a CO image
of KR~140, averaged over the longitude range 133.366$^{\circ}$ to
133.477$^{\circ}$.  Note the ``hole'' at the velocity of the ionized gas
and VES~735.

The question arises as to the relative radial position of VES~735 and
KR~140 with respect to the molecular gas.  In general, it is difficult
to use only the CO radial velocity information to determine absolute
distances to clouds or GMCs.  However, in this direction, Galactic
rotation causes radial velocity to become more negative with increasing
distance.  On the face of it, that would place the molecular cloud on
the far side of the star and nebula.  However, this is possibly too
naive an interpretation because if the cloud is only 10s of pc in radial
extent (like its dimension in the plane of the sky), then the average
shear would be too small to explain the large range in velocity (unless
there were a large enhancement from a density wave).

To address the question of relative position, we made use of both the
\hi and CO data cubes to make an estimate of $A_V$ produced by gas with
velocities out to $-45$~\kms.  As in \S~\ref{sub:co-mass} we used the
conversion factor $X = (1.9\pm0.5) \times 10^{20}$ cm$^{-2}$ (K
km~s$^{-1}$)$^{-1 }$ to convert I(CO) to N(H$_{2}$).  We integrated both
data cubes to obtain the atomic column densities.  At the position of
VES~735 we found N(H\scriptsize I\normalsize) = $4.5 \times 10^{21}$~cm$^{-2}$ 
assuming the emission is optically thin, and N(H$_{2}$) $ = 3.1 \times
10^{21}$~cm$^{-2}$, the latter mostly from local rather than Perseus arm
gas.  Summing the contributions of atomic and molecular hydrogen we
obtain the total hydrogen column density, N$_{H}$ = 2N(H$_{2}$) + 
N(H\scriptsize I\normalsize)$ = 1.1\times10^{22}$~cm$^{-2}$. The total 
visual extinction is computed
using the standard conversion factor, A$_V = 5.3\times10^{-22}$~mag
cm$^{-2}$ (Bohlin, Savage \& Drake 1978). At the position of VES~735 we
obtain A$_V = 5.7 \pm 0.9$, where most of the uncertainty comes from the
uncertainty in the $X$ factor (using the $X$ factors of Digel et al.\
1996 gives a lower $A_V$). This value of $A_V$ compares favorably with
the values around $5.7$ derived by Kerton
\etal\ (1999) using a number of methods (e.g., $B-V$ colour, H$\alpha$
emission measure, DIBs).  However, if the cloud's molecular column
density (summed over $-45$ to $-53$~\kms) of N(H$_{2}$) $ = 2.2 \times
10^{21}$~cm$^{-2}$ (\S~\ref{sub:co-mass}) is included in the extinction
calculation, then the $A_V$ rises to 8.0, which is quite inconsistent
with the observed $A_V$ to VES~735.  We also made maps of the predicted
extinction over the surface of the nebula for comparison with the
extinction map derived by comparing H$\alpha$ surface brightness with
radio emission (Kerton et al.\ 1999).  Again, inclusion of the
extinction from gas in the molecular cloud produces more than two
magnitudes too much extinction.  This is strong evidence that both
VES~735 and the ionized gas of KR~140 lie on the near side of the
molecular cloud gas which has velocities $-45$ to $-53$~\kms.

The extinction measurements combined with Figure~\ref{fig:co-side}
suggest that KR~140 could be a blister \h region on the near side of the
molecular cloud.  This simple geometrical interpretation of KR~140 runs
into difficulty if a systematic champagne flow has developed with gas
flowing away from the parent cloud at up to the sound speed of $~\sim
10$~\kms.  At face value, our H$\alpha$ velocity implies that the
ionized gas is redshifted with respect to the molecular gas, which means
the \h region should be on the far side of the molecular cloud.
However, it can be noted that H$\alpha$ is considered to be a poor line
for estimating the velocity field of champagne flows (Israel 1978; Yorke
\etal\ 1984).  Modeling of line profiles in champagne flows also shows
that the velocity field can be quite random and of low amplitude,
depending on geometry (Yorke \etal\ 1984). Recall that the geometry here
is certainly not plane-parallel and probably more like a broken shell.
Furthermore, our measurement sampled only gas within a projected
distance of 30$^{\prime\prime}$ of VES~735. Fabry-Perot data of the
whole nebula in a line other than a hydrogen line would be a good way to
obtain a better picture of the velocity structure of the ionized gas in
KR~140.

The question then arises as to the origin of the radial velocity spread within
the molecular cloud.  If this is a single, gravitationally bound cloud, 
then this is just the virial velocity, and its mass can be estimated from
\begin{equation}
\label{eq:co-vir-mass}
M_{vir} = {5 R \sigma^2 \over \alpha_{vir} G},
\end{equation}
where $R$ is the length scale for the cloud, $\sigma$ is the linewidth,
and $\alpha_{vir}$ is the virial parameter and includes the effects due
to surface pressure, magnetic fields, and nonuniform densities (Bertoldi
\& McKee 1992).  As in \S~\ref{sub:co-mass}, we take $R=7$~pc, and
measure a typical linewidth to be 5~\kms (see Fig.~\ref{fig:co-side}). 
We take $\alpha_{vir}$ to be
1.1 (Williams \& McKee 1997). With these values $M_{vir} = 1.9 \times
10^5~M_{\odot}$ which is well over an order of magnitude larger than our
measured CO mass of $4.4 \times 10^3~M_{\odot}$.  

This result implies that the velocity field has another origin and can
significantly alter the cloud structure over time.  In fact, recalling
that a velocity of 1~\kms corresponds to 1~pc in a million years, the
``crossing time'', $2 R/\sigma$, is only 3 Myr, comparable to our above
estimates of the age of VES~735 and the \h region.  From the radial
arrangement derived from extinction, it seems that the background
molecular material is moving toward KR 140.  Perhaps the parent cloud
giving rise to VES~735 was the result of a converging flow or ``cloud
collision'' that is still ongoing (relative motion in the plane of the
sky would be expected too).

\section{{\hi} SIGNATURE}
\label{sec:hi}

In theory the presence of \h regions and their surrounding molecular
material should be easily ``visible'' in \hi data sets: in both the
ionized and molecular regions there should be a deficit of \hi emission
due to a lack of neutral atomic hydrogen.  In addition, if the geometry
is favorable, neutral atomic hydrogen associated with the
photodissociation region (PDR) at the ionized-molecular interface should
be visible.  In practice it is actually very difficult to observe \hi
features that are unambiguously associated with an \h region in raw \hi
data sets.  This is primarily due to velocity confusion along the line
of sight caused by the ``turbulent'' motion of the atomic gas.  Since
the velocity of the neutral atomic gas can be many times the typical
channel width in the \hi data sets velocity becomes a much poorer proxy
for physical distance than in a CO data set, which is probing a species
with lower velocity dispersion (heavier, cooler, and less turbulent).
The situation is especially problematic for \h regions in the Galactic
plane where one has to look through a large column of atomic gas towards
the \h region.  For KR~140 we are looking through the local arm and part
way into the Perseus arm of the Galaxy. A preliminary reconnaissance of
the \hi cube confirmed that the complex \hi emission structure along the
line of sight makes seeing any \hi signal associated directly with
KR~140 extremely difficult. Nevertheless, some simple processing of the
\hi data cube does bring out some features associated with the region.

In order to exclude local \hi emission (which is assumed to have a
relatively smooth spatial structure) and to enhance the dynamic range of
the resulting channel maps we constructed a median-subtracted data cube
(Joncas \etal\ 1992; Joncas \etal\ 1985).  In this technique a median
spectrum is calculated for the data cube and then subtracted from each
spectrum making up the cube.  The resulting channel maps thus can
contain both negative and positive values indicating deviations relative
to the median base level.

Figure~\ref{fig:hi-channel} shows channel maps of the median-subtracted
cube over the velocity range $-43.40$ to $-54.95$ km$^{-1}$ bracketing
the ionized gas velocity of $-46\pm2.0$ \kms\ (Kerton \etal\ 1999). We
could not detect any features definitely associated with KR~140 in the
channel maps outside of this velocity range.

Examining these maps we note the following three features.  First, there
is a noticeable deficit of \hi seen in velocity channels $-46.70$ to
$-51.65$ \kms\ outside of the \h region. There is excellent positional
agreement between this deficit in \hi and the observed position of the
CO emission. The deficit can be simply interpreted as being caused by a
lack of \hi emission in the molecular material surrounding
KR~140. Second, we also see a drop in the \hi emission occurring within
the \h region in the $-43.40$ and $-45.05$ \kms\ channels. This deficit
is most likely associated with the ionized gas in KR~140 as suggested by
their spatial correspondence.  Figure~\ref{fig:hispec} presents
spatially averaged \hi and CO spectra for an area just outside of the \h
region to the north-east and the area inside the \h region.  The
anticorrelation between \hi emission and the presence of CO and \h is
evident.  Third, there is an enhancement of \hi emission at
(133.36$^{\circ}$,0.1583$^{\circ}$), seen best in the velocity channel
$-46.70$ \kms. This could be low velocity dispersion material associated
with the PDR. Atomic material in the PDR is expected to have a low
velocity and thus should be seen in channels corresponding to CO
emission.

\section{THE ENVIRONMENT OF KR~140}
\label{sec:discuss}

\subsection{Spontaneous Massive Star Formation}
\label{sub:isolated}

\begin{figure}
\epsscale{1.4}
\plotone{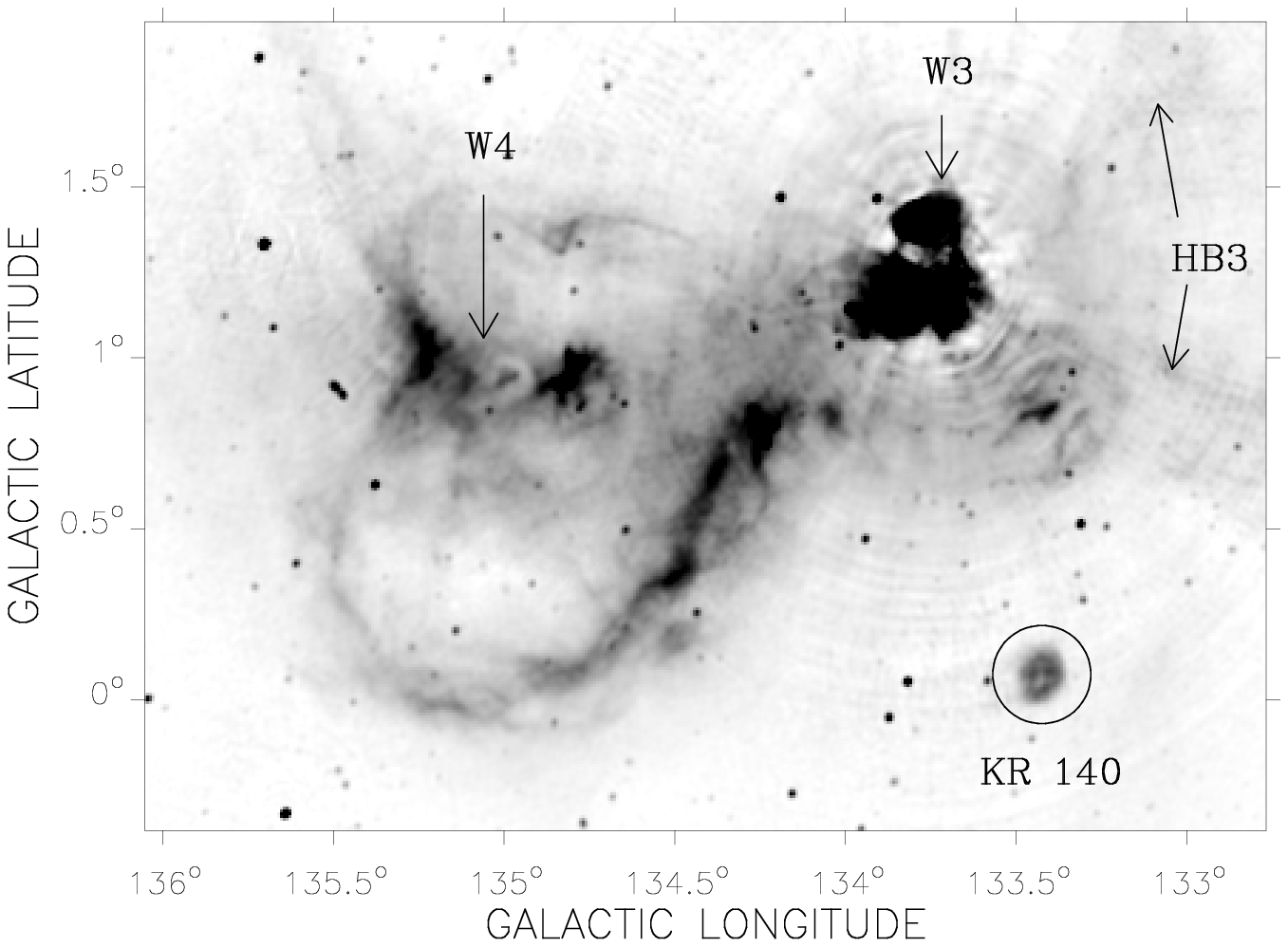}
\caption{KR 140 in context. This 1420 MHz continuum image of the CGPS pilot
region survey shows KR 140 (circled) in relation to the W3/4/5 star forming
regions and the HB3 supernova remnant. The greyscale is linear with a white
value of 2.0~K and a black value of 15.0~K.
\label{fig:1420all}}
\end{figure}

\begin{figure}
\epsscale{1.4}
\plotone{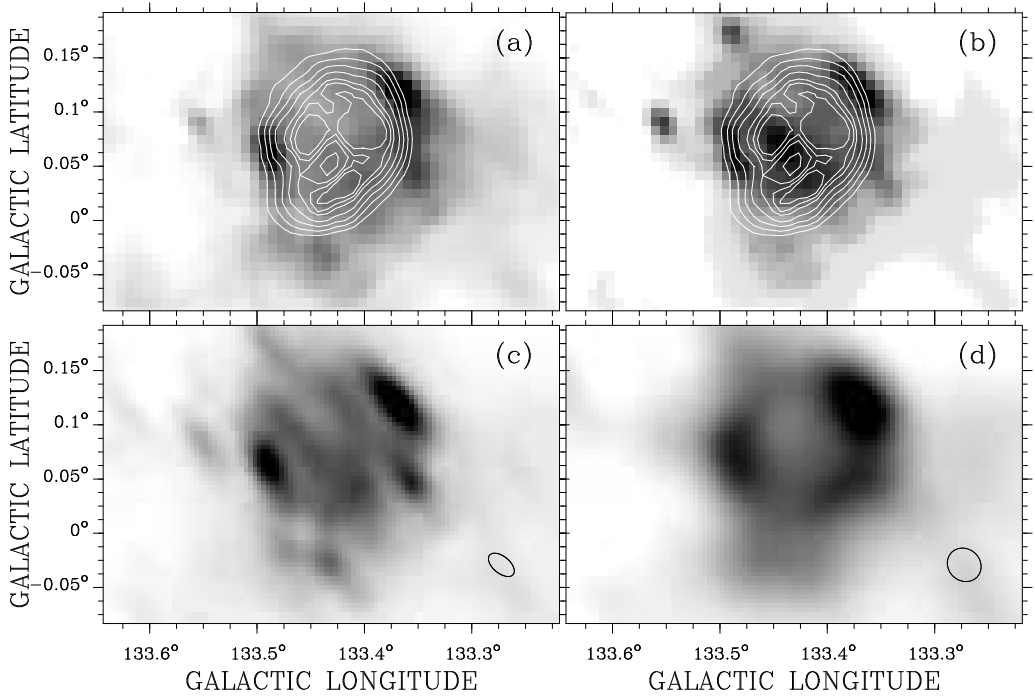}
\caption{{\em (a)} 12 $\mu$m HIRES image convolved to 1420 MHz
resolution, and then overlaid with 1420 MHz contours. The contour values 
are the same as Figure 3{\em a}. The greyscale is linearly scaled from 
2.0 to 20.0~MJy~sr$^{-1}$ (white--black). {\em (b)} 25 $\mu$m HIRES image
convolved to 1420 MHz resolution, and then overlaid with 1420 MHz 
contours. The contour values are the same as Figure 3{\em a} and the image 
is scaled from 4.0 to 25.0~MJy~sr$^{-1}$ (white--black). {\em (c)} Original 
60 $\mu$m HIRES image. The beam shape is shown in the lower right-hand corner. 
The image is scaled from 10.0 to 185.0~MJy~sr$^{-1}$ (white--black). 
{\em (d)} Original 100 $\mu$m HIRES image. The beam shape is displayed in the 
lower right-hand corner. The image is scaled from 40.0 to 350.0~MJy~sr$^{-1}$ 
(white--black). \label{fig:irall}}
\end{figure}

\begin{figure}
\epsscale{1.4}
\plotone{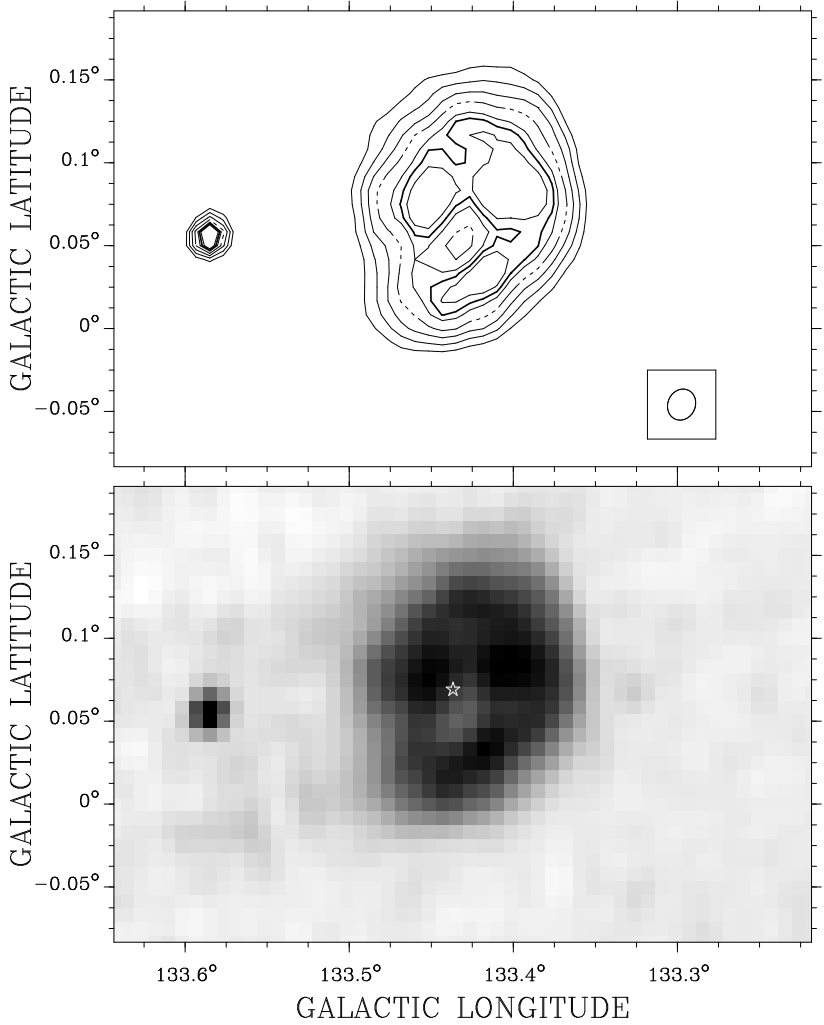}
\caption{(Top) Contours of 21~cm brightness temperature of KR~140 at 5.0, 
6.0, 7.0, 8.0 (dashed), 9.0, 9.5 (heavy), and 10.0~K. The source is slightly 
more extended than indicated by the lowest contour; the \h region fades into 
the background at contour level about 3~K. The beam shape is displayed in the 
lower-right corner. (Bottom) Greyscale with linear brightness scale 
($1.868 - 11.0$~K white--black). The point source at $l = 133.585^\circ$ and 
$b = 0.058^\circ$ is probably extragalactic as it has a non-thermal spectral 
index. The position of the exciting star of this region, VES~735, is denoted 
by the star symbol. \label{fig:kr140}}
\end{figure}

\begin{figure}
\epsscale{0.85}
\plotone{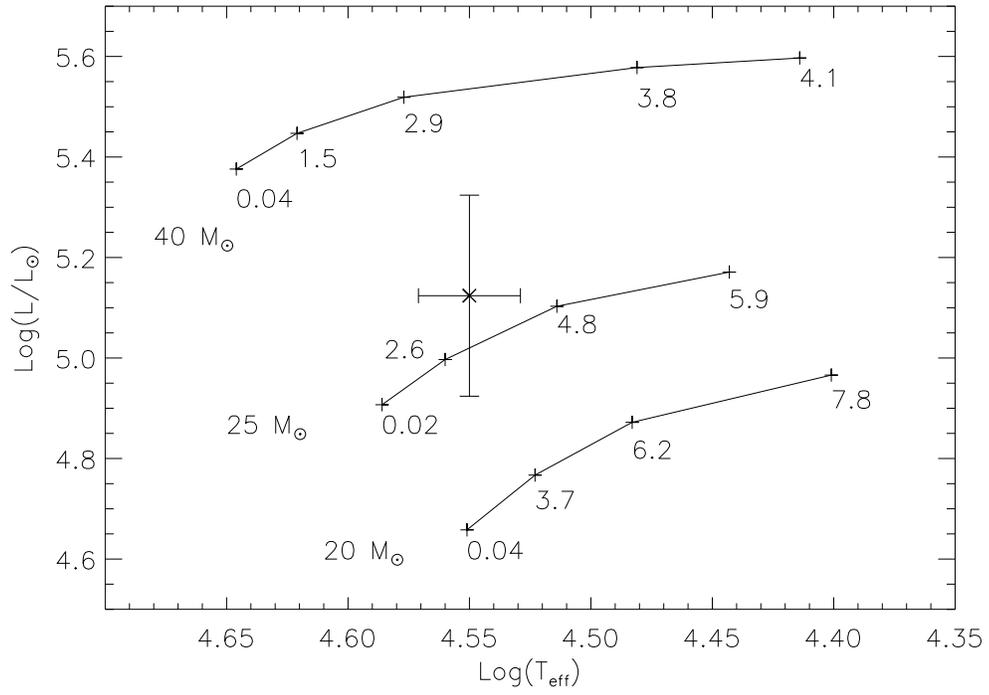}
\caption{$\log(L/L_{\odot})$ vs. $\log T_{eff}$ with evolutionary
models of Schaerer \& de Koter (1997) shown for three high-mass
stars. Individual points in the models (crosses) are labeled with the
stellar age (10$^6$ years).  The point with error bars represents VES
735, the exciting star of KR~140. \label{fig:age}}
\end{figure}

\begin{figure}
\epsscale{0.85}
\plotone{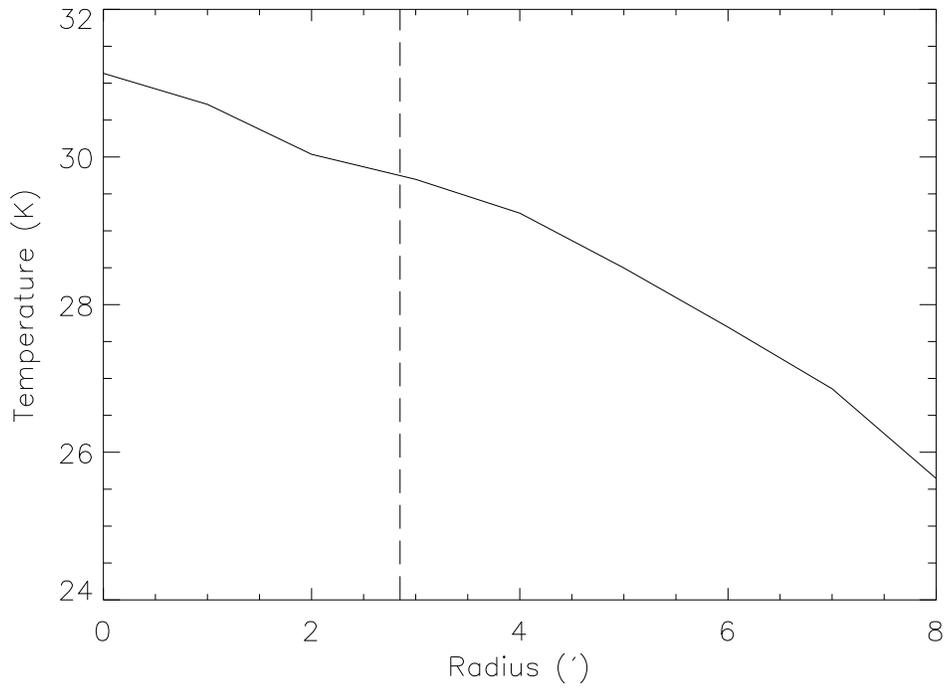}
\caption{Azimuthally averaged dust temperature
profile for KR 140.  Origin is located at $l=133.43^{\circ}$,
$b=0.068$\degr. The extent of the ionized region is indicated by the
dashed vertical line.  The temperature distribution was sampled every
$1'$ in radius and every $10$\degr\ azimuthally.  \label{fig:tdust}}
\end{figure}

\begin{figure}
\epsscale{1.4}
\plotone{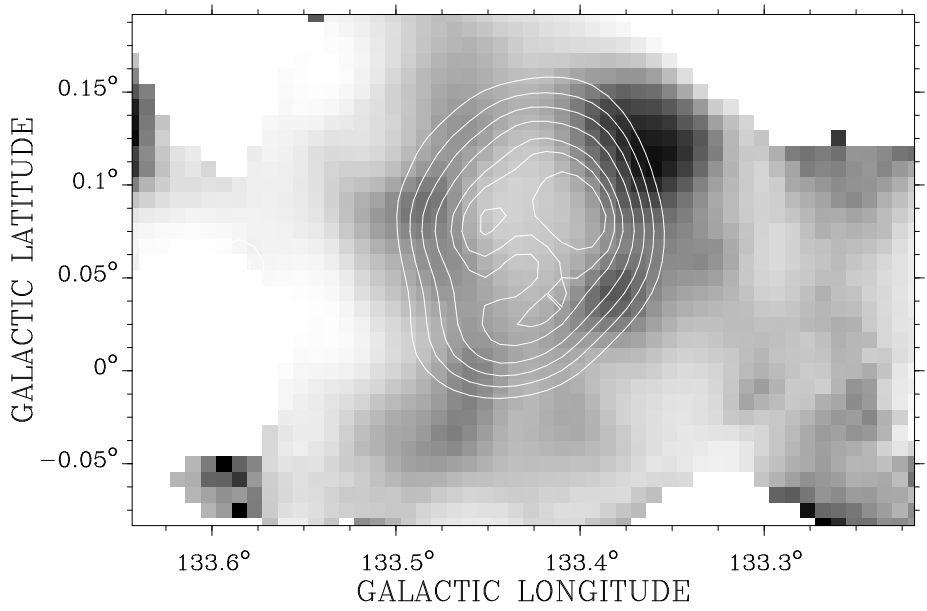}
\caption{$\tau_{100}$ map overlaid with the 1420 MHz contours. The
greyscale is linear with a white value of 0.0 and a black value of
0.002. Both images and the contours are at the 100 $\mu$m resolution. 
\label{fig:tau100}}
\end{figure}

\begin{figure}
\epsscale{1.4}
\plotone{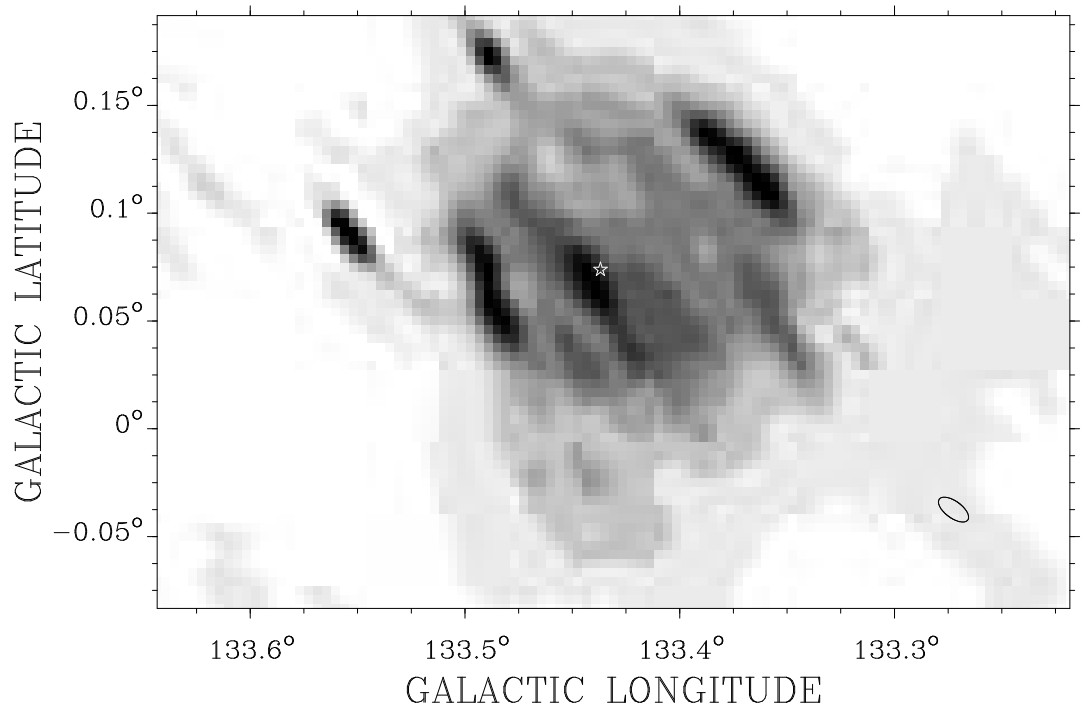}
\caption{The HIRES image of KR 140 at 25$\mu$m. The greyscale 
is linearly spaced with a white value of 4.0~MJy~sr$^{-1}$ and a black
value of 30.0~MJy~sr$^{-1}$. The star symbol near the center of the
nebula corresponds to the position of the ionizing star VES~735.
The beam shape of this HIRES map is shown in the lower right-hand corner.
\label{fig:ir25um}}
\end{figure}

\begin{figure}
\epsscale{0.85}
\plotone{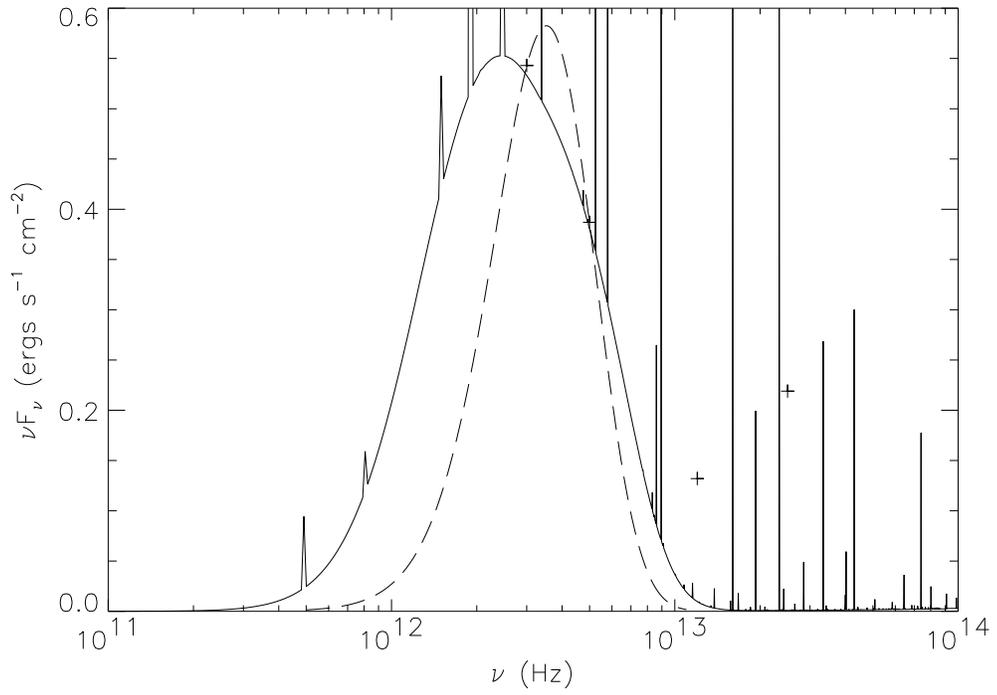}
\caption{A Cloudy model (solid line) using VES 735 stellar parameters
and $f_{c} = 0.44$. Integrated IRAS fluxes for KR 140 are shown by the
crosses.  A single temperature (28.25 K) $\nu^{2}B_\nu$ spectrum is shown
for contrast (dashed line).
\label{fig:cloudy2}}
\end{figure}

\begin{figure}
\epsscale{1.4}
\plotone{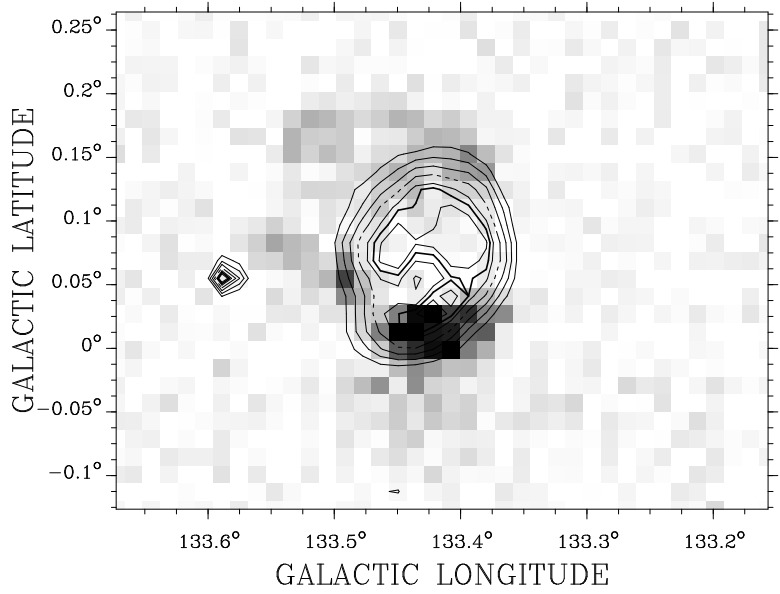}
\caption{An image of integrated CO emission between $-45.53$~km
s$^{-1}$ and $-47.16$~km s$^{-1}$ (LSR). 1420 MHz contours are overlaid
with values that are the same as Figure~\ref{fig:kr140}.
The image is linearly scaled from 0.0 to 15.0~K (white--black).
\label{fig:co-front}}
\end{figure}

\begin{figure}
\epsscale{1.4}
\plotone{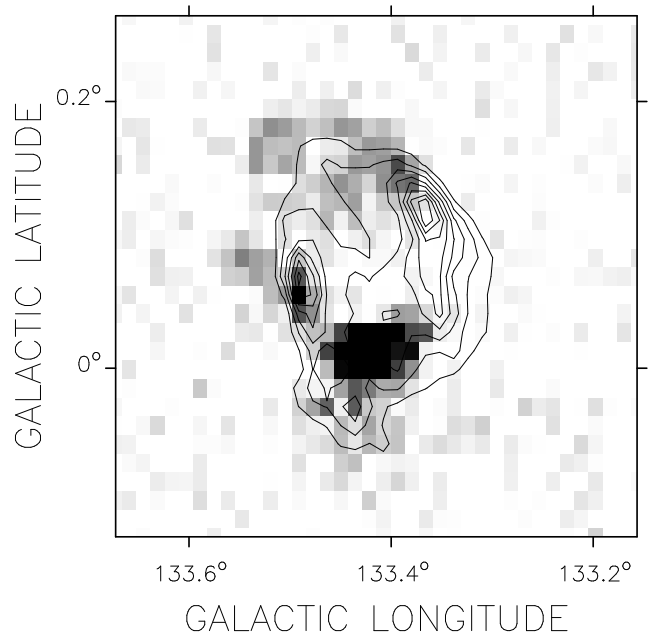}
\caption{Integrated CO emission ($-45.53$ to $-47.16$ (LSR))
associated with KR~140 with contours of HIRES 12~$\mu$m emission, both at
$\sim 1'$ resolution.  The CO image is linearly scaled from 0.5 to
10.0 K (white -- black), the contours are spaced 2 MJy/sr apart from 7.0
to 25.0. \label{fig:co12}}
\end{figure}

\begin{figure}
\epsscale{1.4}
\plotone{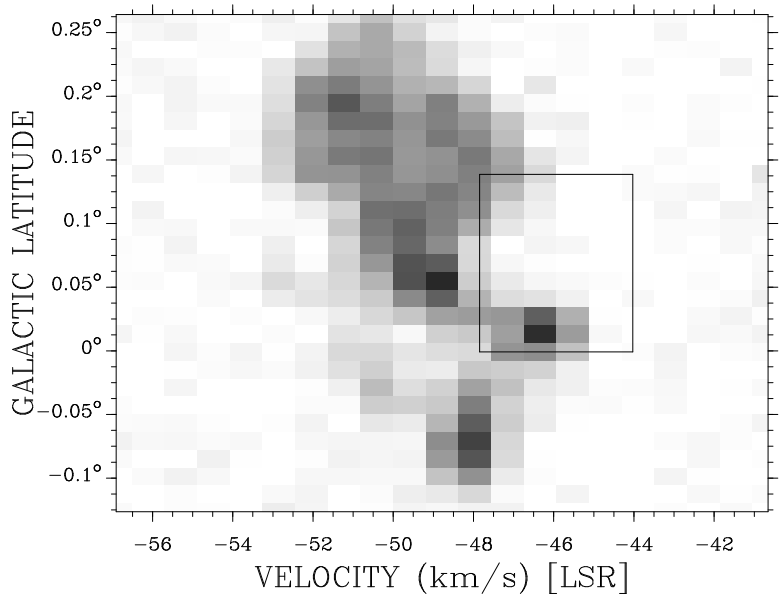}
\caption{Velocity-latitude CO image of KR 140 averaged over the
range $133.366^\circ < l < 133.477^\circ$. The image is linearly scaled
from 0.0 to 7.0~K (white--black). The rectangle denotes the area where
the ionized gas is thought to reside. The latitude range was derived from
the radio continuum image, and the velocity range is derived from 
the measured radial velocity of the ionized gas ($-46 \pm 2.1$~\kms). 
\label{fig:co-side}}
\end{figure}

\begin{figure}
\epsscale{0.95}
\plotone{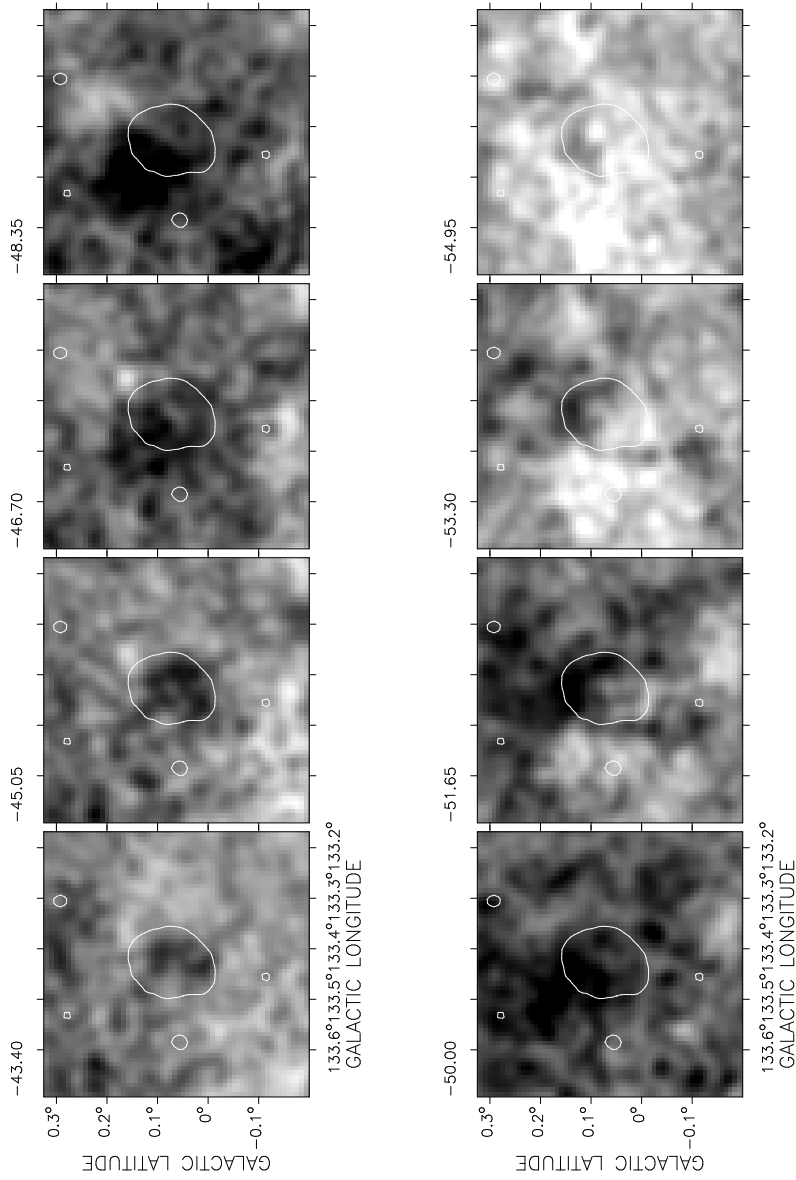}
\caption{\hi channel maps of the region around KR~140 from $-43.40$~\kms\ 
to $-54.95$~\kms.  The \hi maps have been median subtracted and
are linearly scaled from $-2$ to $49$~K (black -- white). The maps have
also been smoothed to 2$^{\prime}$ resolution for clarity. To show the
extent of the ionized  region we have overlaid the 5~K contour from
the 1420~MHz data. \label{fig:hi-channel}}
\end{figure}

\begin{figure}
\plotone{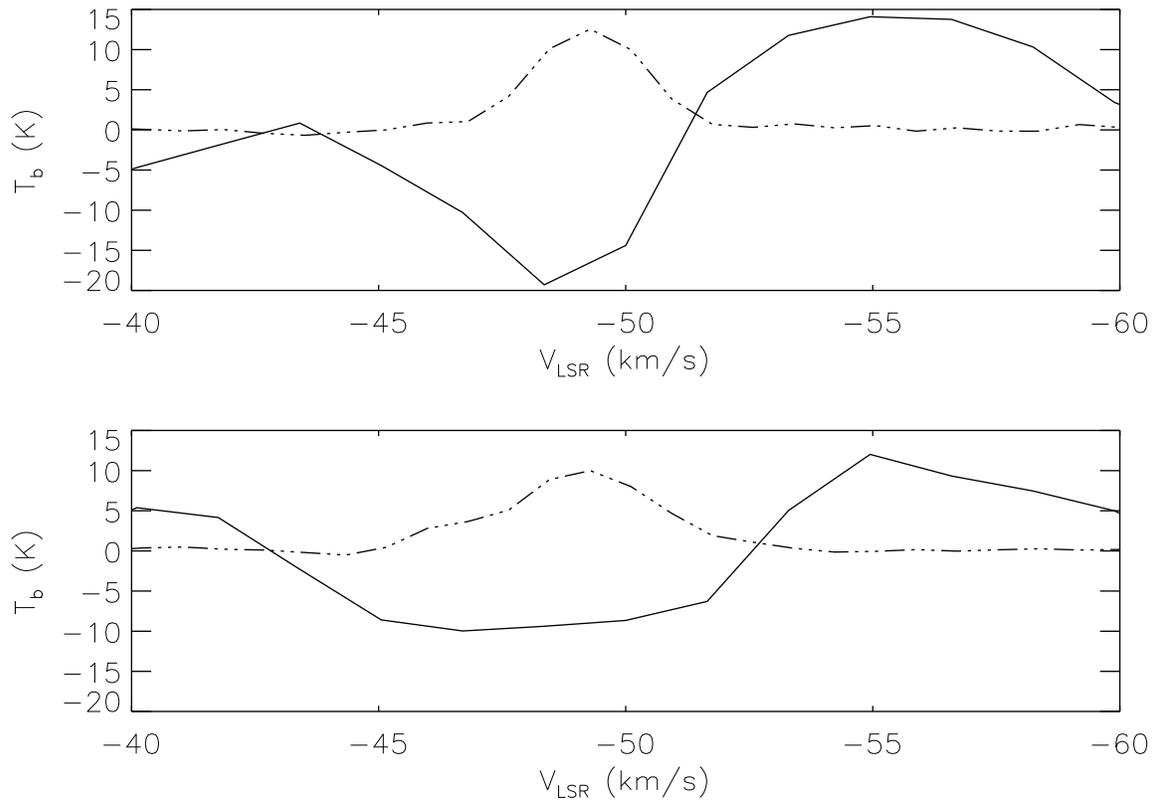}
\figcaption{Spatially averaged CO (dashed line) and \hi (solid line)
spectra for a region containing the CO emission to the northeast of KR~140
 (top), and an area containing the entire \h region (bottom). The CO spectra 
have been multiplied by a factor of two.  The \hi spectra have been median 
subtracted. \label{fig:hispec}}
\end{figure}

In the context of the Perseus Arm star formation activity, the KR~140
complex seems to be unique.  Figure~\ref{fig:1420all} shows that KR~140
is isolated from the massive and violent star formation that is ongoing
around it.  This isolation is evidence to us that KR~140 is an example
of massive star formation in our Galaxy that is \emph{untriggered},
at least in the sense used in the context of sequential star formation.
In none of our data sets does there appear any evidence for a trigger of
the star formation in KR~140.  The exciting cluster of W4, OCl 352, is
about 60~pc away from KR~140 for a cluster distance of 2.35~kpc (Massey,
Johnson \& DeGioia-Eastwood 1995).  For a sound speed of 0.6~\kms
(isothermal speed in H$_2$ at 100~K), the time for a signal to reach
KR~140 would be about 90~Myr, much greater than our estimated age of
KR~140 of a few million years or the age of OCl 352.  From W3, a signal
would take about 60~Myr, but W3 is itself much less than 10$^5$ years
old (Kawamura \& Masson 1998). The supernova remnant HB3 is also in this
complex (Normandeau \etal\ 1997), but from Figure~\ref{fig:1420all}, the
edge of the remnant is nowhere near to KR~140.  Thus, KR~140 does not
seem to be triggered by a neighboring \h region or a nearby supernova
remnant, unless the impulse came along the line of sight.  This
conclusion is complemented by the overall smoothness of the KR~140
nebula; while the nebula does show density inhomogeneities, it is not
far removed from a circular shape.  Thus, any perturbation that might
have triggered the star formation within KR~140 must have been a large
scale phenomenon with a characteristic scale of $\sim 10$~pc. This kind
of triggering might be more consistent with triggering via a spiral
density wave (e.g., Elmegreen 1994, 1995) or by colliding molecular
clouds. Of course, we cannot rule out those kinds of triggers, but the
observational evidence would be difficult to find.

It is interesting to contrast KR~140 to another star forming region that
has been studied with multiwavelength data, the Gemini OB1 molecular
complex (Carpenter, Snell \& Schloerb 1995a, 1995b).  Within the
molecular complex, these authors find young star clusters (from near
infrared data) and a number of dense cores (as identified by CS
observations) associated with IRAS point sources.  Carpenter et al.\
suggest that the arc-shaped morphologies of these cores have been formed
by swept-up gas from expanding \h regions, and that they would form the
massive star clusters in the region.  The other lower infrared
luminosity sources (most likely belonging to lower mass cores) in the
Gem~OB1 complex are not found to be correlated with any arc-shaped
structures or filaments in the molecular gas, and they are not adjacent
to any \h regions; in fact they are spread almost randomly around the
complex.  Carpenter et al.\ therefore conclude that induced star
formation is the prominent mode of formation for massive stars in the
Gem~OB1 molecular complex.

If KR~140 is indeed untriggered, then it seems to be an unusual form
of spontaneous star formation since current observations suggest that the
isolated mode of star formation is generally associated with low mass
star forming regions as seen in Gem~OB1 or even in Taurus. 

\subsection{IRAS Point Sources and Protostars}
\label{sub:ir-point}

\begin{figure}
\epsscale{1.4}
\plotone{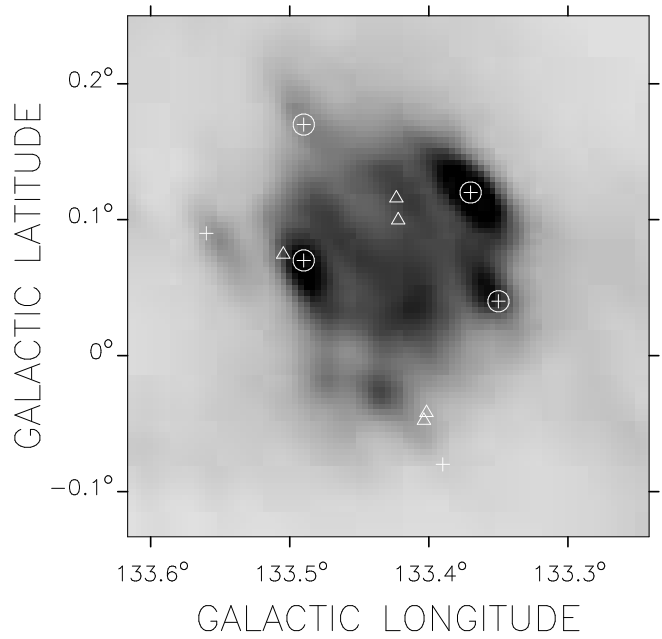}
\caption{IRAS point sources (crosses) and BIRS positions (triangles) near
KR~140. Circled crosses indicate point sources probably associated
with KR~140.  The linear greyscale for the 60~$\mu$m image 
is $0 - 185$~MJy~sr$^{-1}$ (white -- black). \label{fig:psrc}}
\end{figure}

As has been seen in other sites in the Galaxy, an \h region can initiate
star formation via the expansion of its ionization front (e.g.,
Elmegreen \& Lada 1977).  Other stars might also have formed spontaneously.  
Evidence for other star forming regions within
KR~140 may be sought by examining the IRAS point source catalog. In
addition to the IRAS point source 02165+6053 that is cross-referenced to
VES~735, there are six other IRAS point sources in the area in and
around KR~140 (see Table~\ref{tbl:irps}). The crosses in
Figure~\ref{fig:psrc} show the positions of these six IRAS point
sources. The circled crosses are the sources discussed here that seem
most likely to be associated with the KR~140 complex.

The source IRAS~02160+6057 identified with the north-west dust arc
has been the subject of two molecular line investigations. Wouterloot
\& Brand (1989) identified it as a potential star-forming area via its
IRAS colors (see their paper for the exact selection criteria), and
examined it (along with about 1300 other IRAS point sources) for CO
emission. They found a CO feature in that direction at a velocity of
$-$49.7~\kms (LSR), which corresponds to CO in the associated
background molecular cloud, and assigned it the catalog number [WB89]417. This
point source was then observed in an H$_2$O line by Wouterloot, Brand
\& Fiegle (1993), but they were unable to detect any emission. This line
of sight has one one of the highest column densities in the KR~140 \h
region, and so it is possible that a protostar could be forming there as
a result of the expansion of the \h region. However, examination of the
HIRES images shows no point-like features in the dust arc, and none are
found in follow-up submillimetre observations with SCUBA (Kerton \etal\
2000).  Therefore, IRAS~02160+6057 is most likely simply part of the
KR~140 dust shell.  Likewise, IRAS~02168+6052 appears to just be part of
the eastern side of the dust shell; it has not been the subject of any
molecular line observations.

The source IRAS~02157+6053 also seems to have been an identification of
some part of the dust shell.  In the submillimetre there is more
structure in this region; this object appears to be a molecular core
rather than a protostar (Kerton \etal\ 2000).

The point source IRAS~02171+6058 is identified with the dust feature to the
north of the KR~140 complex.  It was included in a CS(2--1) survey by
Bronfman, Nyman \& May (1996) of IRAS point sources that have colors
characteristic of ultracompact \h regions; however, they were unable to
make a detection. Lyder \& Galt (1997) observed this source along with
other ultracompact \h region candidates in a search for methanol maser
emission. Again, they were unable to detect any maser emission from
IRAS~02171+6058. These non-detections do not rule out the possibility
that the source is a protostar.  In fact there is a submillimetre
detection and all evidence seems consistent with a B5V star (Kerton
\etal\ 2000).  It is outside the radio contours of \h region and is
therefore difficult to interpret as being triggered.

A photographic survey of the W3 and W4
region turned up Bright InfraRed Stars (BIRS; Elmegreen 1980) which are
brighter in I band than in R band. There are five of these stars in the
KR~140 region of the sky.  Table~\ref{tbl:birs} gives their coordinates
along with their R and I magnitudes and the positions are indicated by
triangles on Figure~\ref{fig:psrc}.  Elmegreen (1980) estimated that if
these stars were at the distance of the Persues arm then they might be
deeply embedded massive early or pre-main sequence stars, or even giants
or supergiants.  From examining the overall distribution of the BIRS
stars (Figure~1 in Elmegreen 1980), we find there is an overabundance of
BIRS around KR~140 as compared to the average.  It is difficult to say
whether or not these BIRS stars are physically associated with the
KR~140 complex. However, BIRS~128 on the eastern side of the nebula
seems to lie at a well defined edge of the dust shell, and within the CO
shell.

\subsection{The Gas Cloud}
\label{sub:enviro}

We have been able to estimate some properties of the original molecular
cloud which spawned the KR~140 \h region.  Originally having $M \sim 4.9
\times 10^3~M_{\odot}$ this cloud would then be classified as a dwarf
molecular cloud.  These are quite numerous in the ISM. The column
density of $2.2 \times 10^{21}$~cm$^{-2}$ is about the same as the
median column density of $2 \times 10^{21}$~cm$^{-2}$ in the molecular
gas in the outer Galaxy (Heyer \etal\ 1998).  Once VES~735 formed in the
molecular cloud, it had a tremendous impact on the environment around
it.  It has incorporated 25~$M_\odot$, ionized about 160~$M_\odot$, and
a further $290~M_{\odot}$ of cloud material has been removed since the
ionization front broke out of the front part of the molecular cloud.
The remnant cloud is of fairly low density, $n_{H_2} \sim
100$~cm$^{-3}$.

The velocity width of the molecular cloud is too large for the cloud to be
bound, but it is unlikely to arise from velocity shear due to differential
Galactic rotation. It is possible to interpret the velocity structure as a large 
scale flow of material towards KR~140. If true, this flow could have had an impact
on the star formation history of this region.

\begin{figure}
\epsscale{0.85}
\plotone{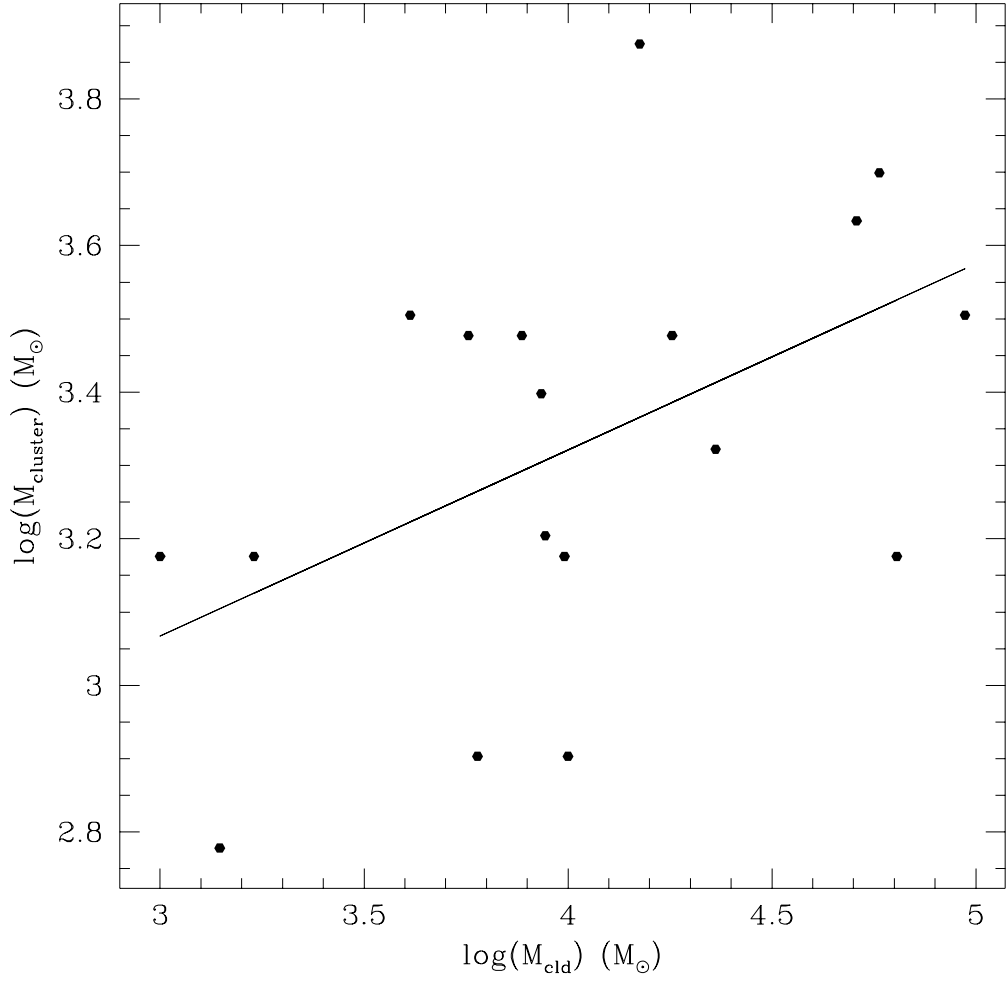}
\figcaption{The correlation between the mass of a molecular cloud and the
mass of the stellar cluster that has formed within the cloud. Derived
from data given by Hunter \etal\ (1990). The rms scatter in the relation is
0.245 dex. \label{fig:hunt}}
\end{figure}

We can use our estimate of the original molecular cloud mass to estimate
the mass of the stellar cluster that likely formed along with
VES~735. Hunter \etal\ (1990) observed clouds in our mass range and
estimated cluster masses using a Miller-Scalo initial mass function
(IMF; Miller \& Scalo 1979).  Using their data we find the correlation
shown in Figure~\ref{fig:hunt}: $\log M_{cluster} = (0.254 \pm 0.111)
\log M_{cld} + (2.306 \pm 0.449)$.  With the above cloud mass we
estimate a cluster mass of about $1.7 \times 10^3~M_{\odot}$.  Of
course, since we know the mass of the most massive star we can use a
Miller-Scalo IMF to estimate the mass of the cluster directly.  Using
Equation~12 in Elmegreen (1983) we estimate a cluster mass of about $1.6
\times 10^{3}~M_{\odot}$, consistent with the above ``empirical''
estimate.  Using these cluster and cloud masses we estimate a star
formation efficiency (SFE = $M_{cluster}/(M_{cluster} + M_{cld})$) of
about 25\% which is typical for clouds of this size (Hunter \etal\
1990), although values can range over two orders of magnitude from cloud
to cloud (Williams \& McKee 1997).  This low value of the SFE implies
that the cluster within KR~140 will not be gravitationally bound.

Near infrared observations (such as ones in the ongoing 2MASS survey)
should be able to detect a number of pre-main sequence stars from this
cluster around VES~735.  It would be interesting to compare the IMF in
this region to the IMF in regions where the star formation was
triggered.  Which, if either, is a good description of the IMF in the
Galaxy?

\section{CONCLUSIONS}
\label{sec:concl}

We have utilized our multiwavelength data set (all at a resolution of
about 1$^{\prime}$) to study not only the physics of the \h region
itself, but also, since the data are from the larger CGPS survey, to
study KR~140 in the context of the overall picture of star formation in
the Perseus spiral arm. We find no evidence for a mechanism that
triggered the formation of the O8.5V(e) star, VES~735, and its (largely
unseen) cluster. We therefore conclude that this region formed
spontaneously out of its parent molecular cloud, independent of the more
vigorous star formation in W3 and W4 nearby.

Our data of KR~140 are consistent with the model of a bowl-shaped region
viewed close to face on.  Extinction measurements to the exciting star,
VES~735, and nebula show that the \h region is quite likely on the near
side of its molecular cloud.  We have not observed any champagne flow,
and cannot rule out other geometries.  KR~140 has an age less than a few
million years.  We have estimated that the original molecular cloud had
a mass of $4.9 \times 10^3~M_{\odot}$ and an average density about
100~cm$^{-3}$, which classifies it as a dwarf molecular cloud. This
makes KR~140 even more unusual as it is a rare example of an O star that
has formed in a cloud with a mass less than 10$^4~M_{\odot}$. 
There is tentative evidence that the molecular material is undergoing
a large scale flow towards KR~140. Follow-up observations are needed 
to pursue this idea.
There are four IRAS point sources associated with the KR~140 complex,
one of which is a possible protostar candidate and another a molecular
core. Near infrared observations of KR~140 are needed to find and study
the young cluster that likely formed along with VES~735.

\acknowledgements

We would like to thank Gary Ferland for his assistance with the use of
Cloudy and Doug Johnstone for helpful discussions.  We also acknowledge
useful comments by the referee, Peter Barnes.  This research made use of 
the SIMBAD data base, operated at CDS, Strasbourg, France. This research 
was supported by the Natural Sciences and Engineering Research Council of 
Canada. D.R.B.\ participated originally via the Physics Co-op program of 
the University of Victoria.

\clearpage

\clearpage

\clearpage

\clearpage

\clearpage

\clearpage

\clearpage

\clearpage

\clearpage

\clearpage

\clearpage

\clearpage

\clearpage

\clearpage

\clearpage

\clearpage

\begin{deluxetable}{lll}
\small
\tablecaption{KR 140 Physical Properties \label{tbl:physprop}}
\tablewidth{0pt}
\tablehead{\colhead{} & \colhead{Property} & \colhead{Value}}
\startdata
General &  Distance\tablenotemark{a}  &  2.3$\pm$0.3 kpc \\
        &  Diameter  &  5.7 pc          \\
Exciting Star\tablenotemark{a} &  Name      &  VES 735  \\
        &  Apparent $V$ Magnitude & $12.88 \pm 0.01$ \\
        &  Apparent $B-V$ & $1.53 \pm 0.02$ \\
        &  Spectral Type & O8.5 V(e) \\
        &  $\log[Q(H^0)]$    &  $\sim 48.5$ \\
        &  $L_{\rm bol}$ &  $\sim 10^5$ $L_\odot$ \\
Radio Cont. (1420 MHz) &  Flux Density  & 2.35$\pm$0.05 Jy \\
        &  Avg. Emission Measure & 2000~cm$^{-6}$~pc \\
        &  $n_e (rms)$   & 30 cm$^{-3}$ \\
        &  $M_{\rm HII}$ & 160 $M_\odot$ \\
        &  $\log[Q(H^0)]$    & 48.05$\pm0.1$ \\
Infrared (IRAS)\tablenotemark{b}&  $F_{12}$ & 110 Jy \\
        &  $F_{25}$ & 140 Jy \\
        &  $F_{60}$ & 970 Jy \\
        &  $F_{100}$ & 2300 Jy \\
        &  $L_{\rm bol}$ & $10^{4.5} L_\odot$ \\
Molecular Line (CO J=1-0)       &  $n_{H_2}$ & $\sim 100$ cm$^{-3}$ \\
        &  Current $M_{\rm total}$ & 4400~$M_\odot$ \\
\enddata
\tablenotetext{a}{Data from Kerton \etal\ (1999)}
\tablenotetext{b}{Fluxes measured from HIRES images of KR 140}
\end{deluxetable}

\clearpage

\begin{deluxetable}{lcccc}
\tablecaption{Radio Flux Factors \label{tbl:radmod}}
\tablewidth{0pt}
\tablehead{\colhead{Model} & \colhead{Abundance\tablenotemark{a}} &
\colhead{Grains\tablenotemark{b}} & \colhead{Temperature} &
\colhead{$\nu F_\nu$ (ergs s$^{-1}$ cm$^{-1}$)}}
\startdata
1 &  \h & Orion &  variable  &  4.2 \\
2 &  \h & ISM   &  variable  &  3.6 \\
3 &  \h & none  &  variable  &  4.6 \\
4 & no He & Orion & variable &  3.9 \\
5 & no He & ISM & variable & 3.3 \\
6 & \h & Orion & fixed (7500K) & 3.3 \\
\enddata
\tablenotetext{a}{\h -- average \h abundance as reported by Ferland (1996)}
\tablenotetext{b}{ISM -- typical high R grains; Orion -- low R grains;
see Ferland (1996) for details}
\end{deluxetable}

\clearpage

\begin{deluxetable}{lcccccccc}
\tablecaption{IRAS Point Sources Near KR 140 \label{tbl:irps}}
\tablewidth{0pt}
\tablehead{\colhead{IRAS} & \colhead{$l$\degr} & \colhead{$b$\degr} &  \colhead{Notes}
}
\startdata
02168+6052 & 133.49  &  \phs0.07 & part of dust shell \\
02160+6057 & 133.37  &  \phs0.12 & part of dust shell \\ 
02157+6053 & 133.35  &  \phs0.04 & part of dust shel; molecular core \\
02171+6058 & 133.49  &  \phs0.17 & no CS or CH$_3$OH detection; colours of UC\h region \\
02174+6052 & 133.56  &  \phs0.09 & not associated (?) \\
02156+6045 & 133.39  &   $-$0.08 & not associated (?) \\
\enddata
\end{deluxetable}

\clearpage

\begin{deluxetable}{lcccc}
\tablecaption{Bright Infrared Stars (BIRS) Near KR 140 \label{tbl:birs}}
\tablewidth{0pt}
\tablehead{\colhead{Star} & \colhead{$l$\degr} & \colhead{$b$\degr} &  \colhead{R}  &  \colhead{I}}
\startdata
BIRS 128 & 133.50472  &  \phs0.07443 & 20.5 & 13.8 \\
BIRS 129 & 133.42336  &  \phs0.11588 & 19.6 & 13.1 \\
BIRS 130 & 133.42205  &  \phs0.09973 & 20.5 & 13.8 \\
BIRS 131 & 133.40169  &   $-$0.04194 & 18.7 & 13.6 \\
BIRS 132  & 133.40358 &   $-$0.04767 & 18.0 & 13.7 \\
\enddata
\end{deluxetable}

\clearpage

\end{document}